\documentclass[aps,nofootinbib,11pt]{revtex4}
\usepackage[letterpaper,hmargin=1in,vmargin=1in]{geometry}

\usepackage{graphicx,epstopdf,amsmath,amsfonts,amssymb,amsxtra,color}

\usepackage{multirow,array}
\usepackage{cancel,bbold,extarrows}
\usepackage{hyperref}

\usepackage{aas_macros} 

\def\be{\begin{equation}}
\def\ee{\end{equation}}
\def\ba{\begin{eqnarray}}
\def\ea{\end{eqnarray}}
\def\ge{\mathrel{\raise.3ex\hbox{$>$\kern-.75em\lower1ex\hbox{$\sim$}}}}
\def\la{\mathrel{\raise.3ex\hbox{$<$\kern-.75em\lower1ex\hbox{$\sim$}}}}

\def\thesection{\arabic{section}}
\def\theequation{\arabic{equation}}
\def\simgt{\mathrel{\raise.3ex\hbox{$>$\kern-.75em\lower1ex\hbox{$\sim$}}}}
\def\simlt{\mathrel{\raise.3ex\hbox{$<$\kern-.75em\lower1ex\hbox{$\sim$}}}}

\newcommand{\nc}{\newcommand}

\nc{\gone}{\bar g_{\pi NN}^{(1)}}
\nc{\gzero}{\bar g_{\pi NN}^{(0)}}
\nc{\al}{\alpha}
\nc{\ga}{\gamma}
\nc{\de}{\delta}
\nc{\ep}{\epsilon}
\nc{\ze}{\zeta}
\nc{\et}{\eta}
\nc{\ka}{\kappa}
\nc{\rh}{\rho}
\nc{\si}{\sigma}
\nc{\ta}{\tau}
\nc{\up}{\upsilon}
\nc{\ph}{\phi}
\nc{\ch}{\chi}
\nc{\ps}{\psi}
\nc{\om}{\omega}
\nc{\Ga}{\Gamma}
\nc{\De}{\Delta}
\nc{\La}{\Lambda}
\nc{\Si}{\Sigma}
\nc{\Up}{\Upsilon}
\nc{\Ph}{\Phi}
\nc{\Ps}{\Psi}
\nc{\Om}{\Omega}
\nc{\ptl}{\partial}
\nc{\del}{\nabla}
\nc{\ov}{\overline}
\nc{\newcaption}[1]{\centerline{\parbox{15cm}{\caption{#1}}}}
\nc{\us}{U(1)$_S$}

\def\beq{\begin{equation}}
\def\eeq{\end{equation}}
\def\bmat{\begin{displaymath}}
\def\emat{\end{displaymath}}
\def\bear{\begin{eqnarray}}
\def\eear{\end{eqnarray}}
\def\ba{\begin{eqnarray}}
\def\ea{\end{eqnarray}}
\def\bery{\begin{array}}
\def\ery{\end{array}}
\def\bit{\begin{itemize}}
\def\eit{\end{itemize}}
\def\ben{\begin{enumerate}}
\def\een{\end{enumerate}}
\def\btab{\begin{tabular}}
\def\etab{\end{tabular}}
\def\btbl{\begin{table}}
\def\etbl{\end{table}}
\def\bfig{\begin{figure}[htb]}
\def\efig{\end{figure}}
\def\bpic{\begin{picture}}
\def\epic{\end{picture}}

\def\ga{\mathrel{\raise.3ex\hbox{$>$\kern-.75em\lower1ex\hbox{$\sim$}}}}
\def\la{\mathrel{\raise.3ex\hbox{$<$\kern-.75em\lower1ex\hbox{$\sim$}}}}
\def\gappeq{\mathrel{\rlap {\raise.5ex\hbox{$>$}}
{\lower.5ex\hbox{$\sim$}}}}
\def\lappeq{\mathrel{\rlap{\raise.5ex\hbox{$<$}}
{\lower.5ex\hbox{$\sim$}}}}

\def\gyr{{\rm \, G\kern-0.125em yr}}
\def\mev{{\rm \, Me\kern-0.125em V}}
\def\gev{{\rm \, Ge\kern-0.125em V}}
\def\tev{{\rm \, Te\kern-0.125em V}}

\begin{document}
 
\title{Kerr-AdS Black Holes and Force-Free Magnetospheres}

\author{Xun Wang and Adam Ritz}
\affiliation{Department of Physics and Astronomy, University of Victoria, 
     Victoria, BC, V8P 5C2 Canada}

\date{February 2014}

\begin{abstract}
\noindent
We obtain analogs of the Blandford-Znajek split monopole solution for force-free magnetospheres around a slowly rotating Kerr-AdS black hole. For small black holes, we find an analytic solution to first order in the ratio of horizon radius to AdS scale, $r_H/l$, which exhibits a radial Poynting flux and for $r_H/l \rightarrow 0$ smoothly approaches the Blandford-Znajek configuration in an asymptotically flat Kerr background. However, for large Kerr-AdS black holes with $r_H/l > 1$, namely those for which the bulk black hole holographically describes the thermodynamics of a strongly-interacting boundary field theory, the existence of a globally well-defined timelike Killing vector external to the horizon suggests the absence of energy extraction through the Blandford-Znajek process. In this regime, we find that at least for slow rotation the force-free solution still exists but exhibits a range of angular velocities for the field lines, corresponding to the freedom in the dual field theory to rotate a magnetic field through a neutral plasma. As a byproduct of this work, we also obtain an analytic solution for a rotating monopole magnetosphere in pure AdS, analogous to the Michel solution in flat space.

\end{abstract}
\maketitle

\section{Introduction}\label{intro}

Rotating Kerr black holes possess an ergosphere, and exhibit the remarkable property that rotational energy can be extracted through purely classical means. The Penrose process, and super-radiance, represent the primary examples. However, through the addition of a force-free magnetosphere, energy extraction through an electromagnetic Poynting flux, via the Blandford-Znajek (BZ) process \cite{BZ} (see also \cite{McKGam,Komissarov:2004ms,GR_GS_eqn,role_ergo_BZ,Gralla:2014yja}), is thought most likely to be realized as a power source in astrophysics, e.g. in active galactic nuclei and quasars. From a theoretical perspective, the force-free magnetosphere induces a perturbative (and possibly a nonperturbative) spin-down of the black hole, as energy is dispersed outward. If the black hole is enclosed in a box, or in anti-de Sitter (AdS) space, where the outgoing modes are reflected back off the boundary then effects such as superradiance are known to lead to a genuine instability. The onset of the superradiant instability in AdS was identified, via the holographic AdS/CFT correspondence \cite{Maldacena:1997re,Witten:1998qj,Gubser:1998bc}, with the limit in which the dual field theory is rotating at the speed of light \cite{Hawking_rotn_AdSCFT}. In this paper, we consider related questions about the BZ process for force-free magnetospheres around Kerr-AdS black holes. 

The implications of embedding the Kerr black hole in AdS depend on the relative size of the black hole, with horizon radius $r_H$, and the AdS curvature scale $l$. For `small' black holes with $r_H\ll l$, the near-horizon geometry is very similar to Kerr, and thus we expect the appearance of an ergosphere and a direct translation of the BZ process as observed in asymptotically flat space. In contrast, for  large black holes with $r_H \geq l$,  the AdS boundary conditions become important and modify the response to the force-free magnetosphere. The Kerr geometry possesses a unique timelike Killing vector as $r\rightarrow \infty$, namely $\xi_{(t)}^\mu$. Using this Killing vector to define energy, one finds an ergosphere outside the horizon, which allows for energy extraction. In contrast, `large' Kerr-AdS black holes possess a family of asymptotically timelike Killing vectors, and thus there is no unique definition of energy for an asymptotic observer at $r\rightarrow \infty$. In the conventional Boyer-Lindquist (BL) coordinate system for Kerr-AdS geometries (with rotation parameter $a$), the angular velocity of zero angular momentum observers (ZAMOs) $\Om_B$, which determines the horizon angular velocity $\Om_H$, is non-vanishing asymptotically where it takes the value $\Omega_\infty=-a/l^2$. Thus, the 
conformal boundary of Kerr-AdS spacetime is an Einstein universe rotating with angular velocity $\Omega_H-\Omega_\infty$. Amongst the family of asymptotically timelike Killing vectors for large black holes, the horizon generator $K_{\Omega_H}^\mu=\xi_{(t)}^\mu+\Omega_H\xi_{(\varphi)}^\mu$ is in fact globally timelike outside the horizon. Moreover, one finds that the boundary Einstein universe rotates slower than the speed of light, provided that $\Omega_H-\Omega_\infty<1/l\Leftrightarrow r_H^2>al$, i.e., for sufficiently large black holes. As argued by Hawking and Reall \cite{HawkingReall}, and discussed below in Section~\ref{BZ_fluxes}, this along with the dominant energy condition (DEC) implies stability of large Kerr-AdS geometries and ensures that energy cannot be extracted. 

Motivated by these arguments, in this work we obtain an analogue of the BZ (split) monopole force-free magnetosphere \cite{BZ}, with the goal of understanding how it evolves  from small to large Kerr-AdS black holes. Recalling that only large black holes, with $r_H > l$, provide saddle points describing the thermodynamics of the holographic dual theory \cite{Hawking_rotn_AdSCFT}, it follows that the stability of the dual thermal state is a direct consequence of the existence of the globally defined timelike Killing vector in the bulk. Although this conclusion suggests the absence of a direct AdS dual of the BZ process, there are at least two interesting subtleties. The first is that stability actually relies on the DEC, which is known to be relatively easy to violate in AdS space, where the Breitenlohner-Freedman (BF) bound allows small negative masses for perturbing fields. Although there is no apparent need for the currents which source the BZ force-free magnetosphere to violate the DEC, this suggests a possible route around the above conclusion that energy extraction is not possible for large AdS black holes. The second subtlety is that the energy defined by the globally timelike Killing vector $K_{\Omega_H}^\mu$ is apparently not the one that naturally enters the thermodynamics of the dual field theory. It has been argued \cite{Gibbons1stlaw} that it is instead the Killing vector $K_{\Omega_\infty}^\mu=\xi_{(t)}^\mu+\Omega_\infty\xi_{(\varphi)}^\mu$ which should be used to define the energy $E$ as use of the conserved charge $E=Q[K_\Omega]$  in the first law $\operatorname{d}E=T\operatorname{d}S+(\Omega_H-\Omega_\infty)\operatorname{d}L$ with $L=-Q[\xi_{(\varphi)}]$ ensures that the r.h.s. is an exact differential. The energy defined in this way does exhibit an ergosphere beyond the horizon even for large Kerr-AdS black holes. This again raises the question of what properties force-free magnetospheres may have for large black holes, given that they should be described in the dual field theory, and motivates finding an explicit bulk solution of this type. 

In their original analysis, Blandford and Znajek obtained an analytic solution in the simplified case where the magnetic field, at zero rotation, is a (split) monopole. While this is an abstraction compared to the physical case where the magnetosphere is induced by an accretion disc, it provides a concrete example in which the radial Poynting flux can be explicitly computed (see also \cite{Menon:2005mg,Brennan:2013jla}). The solution to leading order in the Kerr rotation parameter turns out to be unique, with the axisymmetric magnetosphere co-rotating with a specific angular velocity, equal to half the angular velocity of the horizon. More recent numerical work has confirmed this basic picture (see, e.g., \cite{McKGam}). The primary goal of this paper is to determine the corresponding solution with global AdS boundary conditions. We again work in the slow rotation limit, treating both $a\ll m$ \& $a\ll l$. Away from the small black hole limit, which asymptotically approaches the Kerr case, we find that the field line angular velocity $\omega$ is not uniquely determined. This freedom is also seen to emerge in the AdS analogue to the Michel solution \cite{Michel}, describing a rotating monopole configuration in pure AdS. For large black holes, we interpret these results within the holographic dual in terms of the properties of a fluid in a rotating magnetic field. We find a consistent picture of stable rotation, as the dual fluid is neutral at the corresponding order in the rotation parameter.

The rest of this paper is organized as follows. In Section \ref{gen}, for completeness, we introduce the 3+1 formalism for general stationary and axisymmetric spacetimes, and consider the general features of black hole energy extraction. In Section \ref{setups} we specialize to the Kerr-AdS geometry, and discuss the slow rotation limit. We primarily make use of Boyer-Lindquist (BL) coordinates, while Kerr-Schild (KS) coordinates which are nonsingular on the horizon are discussed in Appendix \ref{Appdx_KS_results}. After these preliminaries, Section \ref{BZsolution} contains the main results: the force-free solution for a rotating monopole in the Kerr-AdS background. In Section \ref{matching} we consider the asymptotic matching of this solution at large radius to a rotating monopole in AdS space. Some implications for the dual field theory are discussed in Section \ref{dualfield}. We finish with some concluding remarks in Section~\ref{conclusion}, including some comments on the membrane paradigm interpretation of the BZ process \cite{mem1,kpm}. The appendices supplement the main text with results presented in Kerr-Schild coordinates, an analytic solution of the BZ force-free magnetosphere for small Kerr-AdS black holes, obtained as an expansion about the BZ solution in the Kerr limit, and lastly a reformulation of the force-free equations in the Newman-Penrose formalism.

\section{Energy extraction from axisymmetric black holes}\label{gen}

In this section, we review the phenomenon of energy extraction from rotating black holes. We present these results in a general form that allows the usual treatment in the Kerr geometry to easily be extended to Kerr-AdS geometries with various coordinate choices.

\subsection{Geometry: \boldmath{$3+1$} formalism}\label{geometry}

We use the $3+1$ formalism \cite{3+1Formalism} which is convenient for presenting our results.
The line element of a general stationary and axisymmetric spacetime can be written as
\begin{equation}\label{metric_lapse_shift}\begin{split}
{\operatorname{d}s}^2&=h_{ij}\bigl(\operatorname{d}x^i+\beta^i\operatorname{d}t\bigr)\bigl(\operatorname{d}x^j+\beta^j\operatorname{d}t\bigr)-\alpha^2{\operatorname{d}t}^2\\
&=h_{ij}\operatorname{d}x^i\operatorname{d}x^j+2\beta_i\operatorname{d}x^i\operatorname{d}t+\bigl(\beta_i\beta^i-\alpha^2\bigr){\operatorname{d}t}^2,\quad (i,j=r,\theta,\varphi).
\end{split}\end{equation}
Given a foliation of the spacetime manifold into constant-$t$ space-like hypersurfaces $\Sigma_t$, \{$h_{ij},\alpha,\beta^i$\} are, respectively, the spatial part of the metric (the projection of $g_{\mu\nu}$ onto $\Sigma_t$), 
the lapse function 
and the shift vector. Geometrically, $h_{ij}$ describes the intrinsic geometry of $\Sigma_t$, $\alpha$ measures the ``distance'' in proper time between two adjacent hypersurfaces $\Sigma_t$ and $\Sigma_{t+\delta t}$ so that $\delta\tau=\alpha\delta t$ and finally $\beta^i$ is the rate of ``shift'' of the coordinate system $\{x^i\}$ on $\Sigma_t$ when evolving in $t$. $\beta^i$ can be seen as part of the 4-vector $\beta^\mu=[\beta^i,0]$, $\beta_\mu=[\beta_i,\beta_i\beta^i]$. In terms of the usual 4-D metric, $h_{ij}=g_{ij},\;\beta_i=g_{it},\;\beta_i\beta^i-\alpha^2=g_{tt},\;(i,j\neq t)$. In this paper we only consider the axisymmetric case with $\beta_\theta(=g_{\theta t})=0$, which is general enough to include both BL and KS forms of the Kerr-AdS metric.

As usual, the geometry admits temporal and azimuthal Killing vectors $\xi_{(t)}^\mu$ \& $\xi_{(\varphi)}^\mu$.
With the present choice of coordinates, these Killing vectors coincide with basis vectors $\partial_t$ \& $\partial_\varphi$. For later convenience, we denote their linear combination by
\begin{equation}
K_\Omega^\mu\equiv\xi_{(t)}^\mu+\Omega\xi_{(\varphi)}^\mu,
\end{equation}
so that $K_\Omega^\mu$ is rotating with angular velocity $\Omega$ relative to $\xi_{(t)}^\mu$. If $\beta^i\ne0$, $\xi_{(t)}^\mu$ fails to be orthogonal to $\Sigma_t$ and can be decomposed as
\begin{equation}\label{xi_t=n+beta}
\xi_{(t)}^\mu=\alpha n^\mu+\beta^\mu,
\end{equation}
where the future-pointing unit normal to $\Sigma_t$ is
\begin{equation}
n_\mu=[0,0,0,-\alpha],\quad n^\mu=\frac{1}{\alpha}[-\beta^i,1].
\end{equation}

The frame-dragging effect in rotating black hole backgrounds is reflected in the fact that an orbiting zero angular momentum observer (ZAMO) has 4-velocity $K_{\Omega_B}^\mu$ with
\begin{equation}
\Omega_B\equiv-\frac{g_{\varphi t}}{g_{\varphi\varphi}}\overset{\text{BL}}{=\joinrel=}-\beta^\varphi,
\end{equation}
where the last equality holds in BL coordinates, which we will work with in the majority of this paper unless otherwise stated. $K_{\Omega_B}^\mu$ is orthogonal to $\Sigma_t$ and the ZAMO is a ``fiducial observer''. Indeed, $K_{\Omega_B}^\mu=\xi_{(t)}^\mu+(-\beta^\varphi)\xi_{(\varphi)}^\mu=\alpha n^\mu$, by \eqref{xi_t=n+beta}. The horizon is where $K_{\Omega_B}^\mu$, as well as $\Sigma_t$, becomes null ($K_{\Omega_B}^\mu\rightarrow K_{\Omega_H}^\mu$, the horizon generator) and is given by $\alpha=0$. On the other hand, $\xi_{(t)}^\mu$ becomes null before the horizon is reached, defining the boundary of the ergosphere (given by $g_{tt}=\beta_i\beta^i-\alpha^2=0$). The horizon can also be viewed as the limiting case of the ZAM-equipotential surfaces $\alpha=\text{const}$.

BL coordinates are singular on the horizon ($g^{tt}=-\alpha^{-2}$, etc.), and thus it is also useful to consider KS coordinates which use a different foliation $\Sigma_{\tilde t}$ that is horizon penetrating. We will make use of BL coordinates for much of the discussion below, as they are analytically more tractable, but the transformation \{$r,\theta,\tilde\varphi(\varphi,r),\tilde t(t,r)$\} to KS coordinates is given in Appendix \ref{Appdx_KS_results}, where we also translate a number of subsequent results for comparison.

\subsection{Black hole energy extraction}\label{energy_extraction}

Using a Killing vector $\xi^\mu$ one can define the conserved \emph{energy-momentum flux vector} $\mathcal{T}^\mu(\xi)\equiv-T^{\mu\nu}\xi_\nu$ (see e.g. \S6.4 of \cite{HawkingWerner_Centenary}), which is non-space-like and future-pointing if $\xi^\mu$ is time-like and future-pointing, due to the dominant energy condition (DEC) \cite{HawkingReall}. Applying Gauss' theorem to $\mathcal{T}^\mu(K_\Omega)$,
\begin{multline}\label{T.xi_Stokes}
0=\int_\mathcal{D}\operatorname{d}^4x\sqrt{-g}\,\mathcal{T}^\mu_{\phantom\mu;\mu}=\int_{\partial\mathcal{D}}\operatorname{d}\mathcal{B}_\mu\mathcal{T}^\mu\\
=\int_{\Sigma_{t_2}-\Sigma_{t_1}}\operatorname{d}^3x\sqrt{^3g}\,n_\mu\mathcal{T}^\mu+\int_{\mathcal{H}=\Sigma_{r_H}}\operatorname{d}\mathcal{B}_\mu\mathcal{T}^\mu+\int_{\Sigma_\infty}\operatorname{d}^3x\sqrt{^3g}\,k_\mu\mathcal{T}^\mu,
\end{multline}
where $\operatorname{d}\mathcal{B}_\mu$ is the volume element restricted to the boundary $\partial\mathcal{D}$ of $\mathcal{D}$. Note that $\partial\mathcal{D}$ consists of two constant-$t$ hypersurfaces $\Sigma_{t_1}$ \& $\Sigma_{t_2}$ ($t_2>t_1$) with normal $n_\mu=[0,0,0,-\alpha]$ and two constant-$r$ hypersurfaces $\Sigma_{r_H}$ \& $\Sigma_\infty$ (which are the horizon $\mathcal{H}$ and the time-like AdS boundary at spatial infinity) with normal $k_\mu=[k_r,0,0,0]$. The integral on $\Sigma_\infty$ can dropped if appropriate boundary conditions are chosen. Eq.~\eqref{T.xi_Stokes} then implies that
\begin{equation}
E(\Sigma_{t_2})-E(\Sigma_{t_1})+F_E^\mathcal{H}=0,
\end{equation}
where $E(\Sigma_t)\equiv-\int_{\Sigma_t}n_\mu\mathcal{T}^\mu(K_\Omega)$ and $F_E^\mathcal{H}\equiv-\int_\mathcal{H}\operatorname{d}\mathcal{B}_\mu\mathcal{T}^\mu(K_\Omega)$ are respectively the total energy on $\Sigma_t$ and the \emph{ingoing} energy flux across the horizon.

To evaluate $F_E^\mathcal{H}$, following \cite{HawkingReall,generalizedPenrose}, one makes use of the ingoing null vector $-k_\mu\sim-\nabla_\mu r\perp\mathcal{H}$ and the null generator ${K_{\Omega_H}}_\mu\overset{\mathcal{H}}{=}\alpha n_\mu=-\alpha^2\nabla_\mu t$ of $\mathcal{H}$ normalized according to $(-k_\mu)K_{\Omega_H}^\mu=-1$ and the decomposition $\mathcal{T}_a\omega^a=-(\mathcal{T}_\mu K_{\Omega_H}^\mu)\omega^1-[\mathcal{T}_\mu(-k^\mu)]\omega^4+\mathcal{T}_2\omega^2+\mathcal{T}_3\omega^3$ in the one-form basis $\{\omega^1=-\omega^k,\omega^2=\omega^\theta,\omega^3=\omega^\varphi,\omega^4=\omega^{K_{\Omega_H}}\}$, to find (using e.g. \cite{Fecko_Diff&Lie_for_phys,HawkingWerner_Centenary,poisson_toolkit,generalizedPenrose}),
\begin{equation}
F_E^\mathcal{H}=-\int_\mathcal{H}\mathcal{T}_4{^\star}\omega^1=\int_\mathcal{H}(-\mathcal{T}_\mu K_{\Omega_H}^\mu)(\omega^2\wedge\omega^3\wedge\omega^4)=-\Delta t\int_{\mathcal{H}\cap\Sigma_t}\operatorname{d}S\mathcal{T}_\mu K_{\Omega_H}^\mu
\end{equation}
where $\eta_{14}=\eta_{41}=-\eta_{22}=-\eta_{33}=-1$, $\epsilon_{r\theta\varphi t}=1=-\epsilon_{1234}$, $\Delta t$ is simply the time interval, and the last integral is on the 2-D spatial section of the horizon. Finally, note that on $\mathcal{H}$ \cite{Grav_foundn_frontier},
\begin{equation}
n_\mu=\frac{1}{\alpha}{K_{\Omega_H}}_\mu=-\frac{1}{2\kappa\alpha}\nabla_\mu(K_{\Omega_H}^\nu{K_{\Omega_H}}_\nu)=\frac{1}{2\kappa\alpha}\nabla_\mu[f_g(r,\theta)\Delta_r]=\frac{f_g(r,\theta)}{2\kappa\alpha}\nabla_\mu\Delta_r\sim k_\mu.
\end{equation}
It follows that ${K_{\Omega_H}}_\mu$ is in the $\nabla_\mu r$ direction ($\parallel k_\mu$), which is also the $-\nabla_\mu t$ direction ($\parallel n_\mu$), where $\kappa$ is the surface gravity and $f_g(r,\theta)$ is a function of metric components, so we have
\begin{equation}
F_E^\mathcal{H}\propto-\int\mathcal{T}^r(K_\Omega).
\end{equation}

Energy extraction is possible if $F_E^\mathcal{H}\propto-\int\mathcal{T}_\mu(K_\Omega)K_{\Omega_H}^\mu<0$, which implies, given that $K_{\Omega_H}^\mu$ is null on the horizon, that $\mathcal{T}^\mu(K_\Omega)$ must be space-like on (and, by continuity, just outside) the horizon. This in turn implies that the Killing vector $K_\Omega^\mu$ with which $\mathcal{T}^\mu(K_\Omega)$ is defined fails to be time-like in the neighbourhood of the horizon, i.e., the existence of an ergosphere.

Arbitrarily close to the horizon, $K_{\Omega_H}^\mu$ is time-like, meaning that the following inequality always holds on the horizon
\begin{equation}\label{horizn_ineq}
-\mathcal{T}^r(K_{\Omega_H})\geq0,
\end{equation}
so that a local observer co-rotating with $K_{\Omega_H}^\mu$ sees an ingoing energy flux. For an asymptotic observer, on the other hand, who defines energy with $K_{\Omega}^\mu$, \eqref{horizn_ineq} implies
\begin{align}
-\mathcal{T}^r(K_\Omega)&-(\Omega_H-\Omega)\mathcal{T}^r(\xi_{(\varphi)})\geq0\label{fluxes_inequality}\\
\Rightarrow\quad&F_E^\mathcal{H}-(\Omega_H-\Omega)L_E^\mathcal{H}\geq0\\
\Rightarrow\quad&\delta E-(\Omega_H-\Omega)\delta L\equiv T\delta S\geq0,\label{1st2nd_law}
\end{align}
where $L_E^\mathcal{H}\equiv\int_{\mathcal{H}}\operatorname{d}\mathcal{B}_\mu\mathcal{T}^\mu(\xi_{(\varphi)})\propto\int\mathcal{T}^r(\xi_{(\varphi)})$ is the total ingoing angular momentum flux across the horizon, and the last step shows that the derivation leads to the 1st and 2nd laws of black hole thermodynamics. To respect the 2nd law, there must also be an angular momentum extraction ($\delta L<0$) accompanying any energy extraction from the black hole (for $\Omega_H-\Omega>0$, e.g., the Kerr case). For large AdS black holes where we can choose $\Omega=\Omega_H$ energy extraction is absent \cite{HawkingReall}. Similar conclusions follow for the super-radiance process \cite{superrad_KAdS}.

\begin{figure}[tbp]
\centering\includegraphics[scale=0.3]{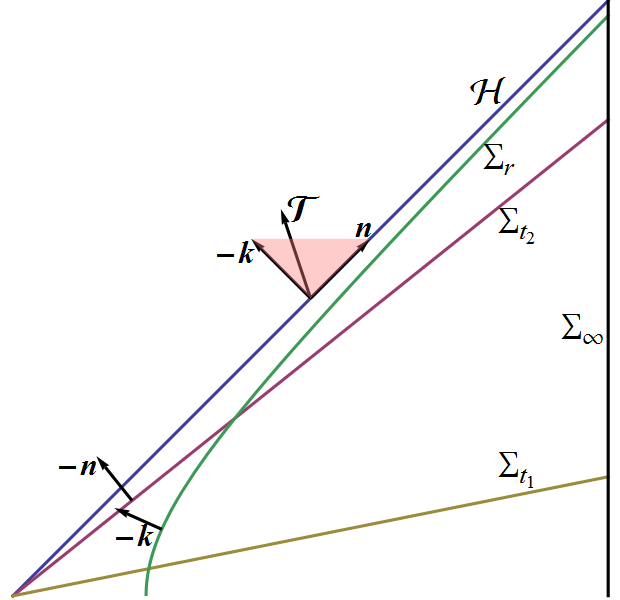}\caption{A spacetime region with horizon $\mathcal{H}$, AdS boundary $\Sigma_\infty$ and constant-$r$(-$t$) slices $\Sigma_{r,t_{1,2}}$ in BL coordinates. With $-n_\mu=\alpha\nabla_\mu t$ and $k_\mu=k_r\nabla_\mu r$, then on $\mathcal{H}$ both ${K_{\Omega_H}}_\mu\sim n_\mu\sim\nabla_\mu r$ and $k_\mu$ are null normals and $\{n_\mu\operatorname{d}x^\mu,k_\mu\operatorname{d}x^\mu,\operatorname{d}\theta,\operatorname{d}\varphi\}$ forms a complete basis.}\label{slices}
\end{figure}

\section{Kerr-AdS and the slow rotation limit}\label{setups}

\subsection{Kerr-AdS solution}
The Kerr-AdS metric in BL coordinates is explicitly given by \cite{Hawking_rotn_AdSCFT}
\begin{equation}
\operatorname ds^2 = - \frac{\De_r}{\Sigma} [\operatorname dt-\frac{a}{\Xi} \sin^2\theta\operatorname d\varphi]^2 + \frac{\Si}{\De_r}\operatorname dr^2 + \frac{\Si}{\De_\theta}\operatorname d\theta^2 + \frac{\De_\theta \sin^2\theta}{\Si} [a\operatorname dt - \frac{r^2+a^2}{\Xi}\operatorname d\varphi]^2.
\end{equation}
where
\begin{align}
\Sigma=r^{2}+a^{2}\cos^{2}\theta, \qquad &\Xi =1-\frac{a^2}{l^2}\\
\Delta_r=(r^{2}+a^{2})\bigl(1+\frac{r^2}{l^2}\bigr)-2mr,& \qquad
\Delta_\theta=1-\frac{a^2}{l^2}\cos^{2}\theta.
\end{align}
In terms of the general axisymmetric metric discussed in the previous section, we have
\begin{align}
h_{rr}&=\frac{\Sigma}{\Delta_r},\qquad h_{\theta\theta}=\frac{\Sigma}{\Delta_\theta},\qquad h_{\varphi\varphi}=\frac{\Delta_\theta(r^{2}+a^{2})^{2}-\Delta_ra^{2}\sin^{2}\theta}{\Xi^{2}\Sigma}\sin^{2}\theta\\
\alpha^{2}&=\frac{\Sigma\Delta_r\Delta_\theta}{\Delta_\theta(r^{2}+a^{2})^{2}-\Delta_ra^{2}\sin^{2}\theta}, \qquad
\beta^i=\Bigl[0,0,-a\Xi\frac{\Delta_\theta(r^{2}+a^{2})-\Delta_r}
{\Delta_\theta(r^{2}+a^{2})^{2}-\Delta_ra^{2}\sin^{2}\theta}\Bigr]. \label{beta_rot}
\end{align}

It is convenient to define the following dimensionless ratios,
\be
 \xi \equiv \frac{a^2}{l^2}, \qquad x \equiv \frac{r_H^2}{l^2},
\ee
where the horizon radius $r_H$ is the largest root of $\Delta_r=0$, which can be written in the form
\begin{equation}
\label{r1rH_eqn}
\De_r(r_H)=0 \;\; \Longleftrightarrow \;\; \frac{l}{2m}(x+\xi)(x+1)=\sqrt{x}.
\end{equation}
This relation is useful for analyzing the various limiting cases as discussed below.
The angular velocity $\beta^\varphi$ in (\ref{beta_rot}) varies from the horizon,
\be
\Omega_H=-\beta^\varphi|_{r_H} = \frac{a\Xi}{r_H^2+a^2},
\ee
to the boundary,
\be
\Omega_\infty = -\beta^\varphi|_{\infty} = -\frac{a}{l^2},
\ee
and this feature of BL coordinates has to be taken into account in considering the reference frame of the holographic dual theory.

For the discussion of slow rotation below, it is also convenient to define the critical mass parameter \cite{susyAdSBHs},
\be
m_{\text{ext}}(a)=l\frac{\bigl[\bigl((1+\xi)^2+12\xi\bigr)^{1/2}+2(1+\xi)\bigr]\bigl[\bigl((1+\xi)^2+12\xi\bigr)^{1/2}-(1+\xi)\bigr]^{1/2}}{3\sqrt{6}}, \label{mext}
\ee
and the constraint $m \geq m_{\rm ext}$ ensures the absence of naked singularities, with $m=m_{\rm ext}$ being the extremal Kerr-AdS black hole with $\De_r$ having a double root at the degenerate horizon. The Killing vectors $\xi_{(t)}^\mu$ and $\xi_{(\varphi)}^\mu$ can be used to define mass and angular momentum (e.g., through Komar formulae in asymptotic flat spacetimes and the conformal definition \cite{Gibbons1stlaw} in asymptotic AdS spacetimes), but the ambiguity in the asymptotic timelike Killing vector noted in Section~\ref{intro} implies that the definition of energy is not unique. This is significant for the analysis of energy extraction, and we discuss the range of asymptotically timelike Killing vectors in more detail below.

\begin{figure}[tbp]
\centering\includegraphics[scale=0.64]{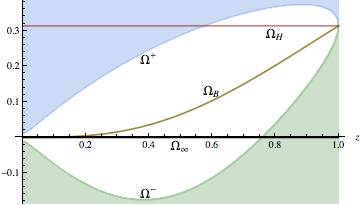}\hspace*{0.2cm}\includegraphics[scale=0.64]{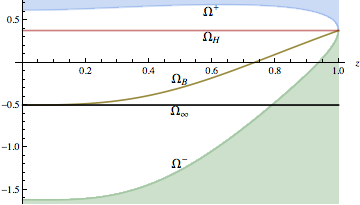}
\caption{The condition $\Omega^-\leq\Omega\leq\Omega^+$ for $K_\Omega^\mu$ to be non-space-like is shown for small (left) and large (right) Kerr-AdS black holes, where $z\equiv\frac{r_H}{r}=1$ and $0$ are the horizon and boundary at infinity. The ergoshere for $K_\Omega^\mu$ is where the horizontal line $\Omega=\text{const.}$ is within the shaded regions. The lines show specific values of angular velocity, $\Omega=\Omega_H$ and $\Omega=\Omega_\infty$, plus the axis $\Omega=0$. $\Omega_B(z=1)=\Omega_H$ and $\Omega_B(z=0)=\Omega_\infty=-\frac{a}{l^2}$.
Left: $\{m=1,a=0.9,l=100,\cos\theta=0.5\}$ for a faster rotating and smaller black hole, close to the Kerr limit. $\Omega_\infty$ and the axis are indistinguishable. There is no globally time-like Killing vector $K_\Omega^\mu$.\\
Right: $\{m=1,a=0.5,l=1,\cos\theta=0.5\}$ for a slower rotating and larger black hole. $K_{\Omega_H}^\mu$ is now globally time-like.}\label{Omega_range}
\end{figure}

The general condition for a Killing vector $K_\Omega^\mu\equiv\xi_{(t)}^\mu+\Omega\xi_{(\varphi)}^\mu$ to be non-space-like requires $\Omega^-\leq\Omega\leq\Omega^+$ where 
\begin{equation}
\Omega^\pm =\Omega_B\pm\sqrt{-\frac{K_\Omega^\mu{K_\Omega}_\mu}{h_{\varphi\varphi}}} \overset{\text{BL}}{=\joinrel=}\Omega_B\pm\frac{\alpha}{\sqrt{h_{\varphi\varphi}}}.
\end{equation}
These bounding contours are shown for two cases in Fig.~\ref{Omega_range}. 

The conformal boundary of Kerr-AdS spacetime is an Einstein universe rotating with angular velocity $\Omega = \Omega_H-\Omega_\infty$, where $\Omega_\infty=-a/l^2$ is the angular velocity of the non-rotating frame at infinity. A feature of the Kerr-AdS geometry is that, as seen from the plot, the Killing vector $K_{\Omega_H}^\mu=\xi_{(t)}^\mu+\Omega_H\xi_{(\varphi)}^\mu$ is globally time-like, and the Einstein universe rotates slower than the speed of light, provided $\Omega_H-\Omega_\infty<1/l\Leftrightarrow r_H^2>al$; i.e., for large black holes \cite{Hawking_rotn_AdSCFT,Thermo_KerrAdS_CFT,GibbonsErgoMagBH}. 
The critical angular velocity for the Einstein universe to rotate at the speed of light corresponds to $\Omega_H=\Omega^+(r\rightarrow\infty,\theta=\pi/2)=\Omega_\infty+1/l$. Usually the ergosphere for $K_\Omega^\mu$ starts at $\Omega=\Omega^-$ and extends to the horizon \cite{HawkingEllis}. The existence of an ergosphere is of course essential for any energy extraction mechanism from black hole. However, the plot reveals that there is no unique time-like Killing vector $K_\Omega^\mu$ at infinity, hence the ambiguity in defining energy. As argued by \cite{HawkingReall} and shown explicitly for the BZ process in Section \ref{BZ_fluxes}, with energy defined with the globally time-like $K_{\Omega_H}^\mu$ there is no energy extraction and the black hole is stable. On the other hand, consideration of the thermodynamics of the dual field theory \cite{Gibbons1stlaw} suggests that $K_{\Omega_\infty}^\mu$ is the appropriate choice of Killing vector to use in defining energy; namely the unique choice that yields the first law $\operatorname{d}E=T\operatorname{d}S+(\Omega_H-\Omega_\infty)\operatorname{d}L$ with the r.h.s. an exact differential. Here we are adopting the definitions of energy and angular momentum as conserved charges associated with Killing vectors, denoted $E=Q[K_\Omega]$ and $L=-Q[\xi_{(\varphi)}]$ \cite{Gibbons1stlaw}, with $Q[\xi_{(t)}]=\frac{m}{\Xi}$ and $L=\frac{ma}{\Xi^2}$ for Kerr-AdS.\footnote{Note that, without spoiling the exactness of the r.h.s., one can in principal choose a different $K_{\Omega'}^\mu$ with $\Delta\Omega=\Omega'-\Omega_\infty$ independent of $\{L,S\}$, so that $\operatorname{d}(E-\Delta\Omega L)=T\operatorname{d}S+(\Omega_H-\Omega_\infty-\Delta\Omega)\operatorname{d}L$. In terms of the independent variables $\{L,S\}$, we have,
$r_H^2= \frac{S}{\pi}\biggl[\dfrac{4L^2}{(S/\pi)^2(S/(l^2\pi)+1)}+1\biggr]^{-1}$ and $a^2=\frac{S}{\pi}\biggl[\dfrac{(S/\pi)^2(S/(l^2\pi)+1)^2}{4L^2}+\frac{S}{l^2\pi}+1\biggr]^{-1}$.}
The ambiguity in the definition of energy motivates a more detailed investigation of the BZ process and force-free magnetospheres even for large Kerr-AdS black holes. We will turn to this topic in the next section, after describing some useful features of the slow rotation limit.

\subsection{Slow rotation}

The Kerr-AdS solution is characterized by three parameters $\{m,a,l\}$ or equivalently $\{r_H,a,l\}$. Slow rotation generically implies a regime far from extremality, set by $m\gg m_{\text{ext}}$ (see \eqref{mext}) or alternatively $r_H\geqslant r_H^\text{ext}$, where
\begin{equation}
r_H^\text{ext}=l\biggl[\frac{-(1+\xi)+\bigl((1+\xi)^2+12\xi\bigr)^{1/2}}{6}\biggr]^{1/2}.
\end{equation}
$m_{\text{ext}}$ \& $r_H^\text{ext}$ are the extremal limits of $m$ \& $r_H$ (see e.g. \cite{KAdS/CFTdiverseD_m&rH_extr}). For Kerr ($l\rightarrow\infty,\xi\rightarrow0$), $m_{\text{ext}}=r_H^\text{ext}=a$ and the condition $\frac{a}{m}\ll1$ used in the perturbative solution guarantees $m\gg m_{\text{ext}}$. For Kerr-AdS, $m\gg m_{\text{ext}}$ implies,
\begin{equation}\label{a/m<<a/m_extr}
\frac{a}{m}\ll\frac{a}{m_\text{ext}} 
   \sim {\cal O}(1)
\end{equation}
where the latter condition holds for $\xi \in [0,1]$, and thus $\frac{a}{m}\ll1$ is still a good criterion for ``far from extremality''.

\begin{figure}[tbp]
\centering\includegraphics[scale=0.5]{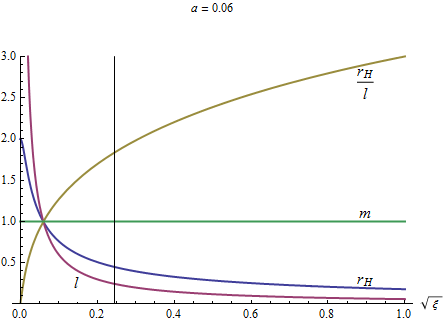}\includegraphics[scale=0.5]{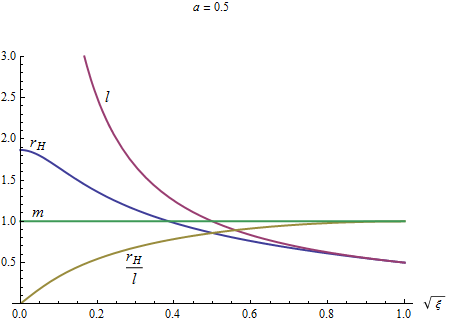}
\caption{$r_H$, $l$ and $\frac{r_H}{l}$ as functions of $\xi$, in units $m=1$. Real solutions for $r_H$ do not exist for all $\xi$ when $\frac{a}{m}\rightarrow1$, as expected. Our small `$a$' expansion is valid for small values of $\xi$ up to the vertical line in the first graph.}\label{sizes}
\end{figure}

The AdS length scale $l$ enables us to talk about black hole sizes in terms of $x\equiv\frac{r_H^2}{l^2}$. To gain some  intuition, we plot $r_H$, $l$ and $\frac{r_H}{l}$ for various $\frac{a}{m}$ in Fig. \ref{sizes}, which shows that large black holes ($x>1$) are only possible for $\frac{a}{m}<\frac{1}{2}$. Indeed, rearranging $\Delta_r(r_H)=0$ formally into a quadratic equation for $r_H$ with fixed $x$: $r_H^2-\frac{2m}{1+x}r_H+a^2=0$, the condition that $r_H$ be real is $\frac{a}{m}<\frac{1}{1+x^2}$. As discussed above, large black holes satisfying $r_H^2>al\Leftrightarrow x>\sqrt{\xi}$ are generically stable \cite{HawkingReall}, so combining the two inequalities we have,
\begin{equation}
\frac{a}{m}<\frac{1}{1+x^2}<\frac{1}{1+\xi},
\end{equation}
with $\frac{1}{1+x^2}\in[0,1]$ and $\frac{1}{1+\xi}\in[\frac{1}{2},1]$, for $l\in(0,\infty)$ keeping $\xi\leq1$. It is then clear that large black holes ($\frac{1}{1+x^2}<\frac{1}{2}$) imply slow rotation and ensure stability, while fast rotation ($\frac{a}{m}>\frac{1}{2}$) allows for an instability (violating the second inequality by insisting on the first one).

In addition to the slow rotation condition $\frac{a}{m}\ll1$, our small `$a$' expansion also treats $\frac{a}{l}\ll1$ and includes the following three regimes according to the relative scale between $m$ and $l$:
\begin{enumerate}
\item $a\ll m\sim\frac{r_H}{2}\ll l$,\quad\text{(small black holes)}\label{a<<m<<l}
\item $a\ll m\sim r_H\sim l$,\quad\text{(intermediate black holes)}\label{a<<m=l}
\item $a\ll l\ll r_H\ll m$,\quad\text{(large black holes)}\label{a<<l<<m}
\end{enumerate}
Note that the criterion $r_H^2>al$ for globally time-like $K_{\Omega_H}^\mu$ is met in regimes \ref{a<<m=l} and \ref{a<<l<<m}, but may or may not be met in regime \ref{a<<m<<l}.
In Fig. \ref{sizes}, regime \ref{a<<m<<l} corresponds to the leftmost region of the graph, regime \ref{a<<m=l} is around the transition point from small to large black holes where the curves meet and regime \ref{a<<l<<m} is further to the right up to the vertical line. To be more precise, in terms of the small parameter $\frac{a}{m}\equiv\epsilon$, the transition point is at $\sqrt{\xi}\sim\epsilon$ and the vertical line bounding regime \ref{a<<l<<m} would be e.g. at $\sqrt{\xi}\sim\epsilon^{\frac{1}{2}}$. The regime to the right of the vertical line is where $\frac{a}{l}\sim\mathcal{O}(1)$ and is not covered by our small `$a$' expansion. It describes very large black hole with sizes diverging as, e.g., $\frac{r_H}{l}\sim\epsilon^{-\frac{1}{3}}$ at $\sqrt{\xi}=1$.

\subsection{Small `$a$' expansion}\label{notation_setup}
To implement the slow rotation limit, it is useful to define
\begin{equation}\label{Delta0}
\Delta_0\equiv\Delta_r(a=0)=\frac{r}{l^2}(r^3+l^2r-2ml^2)=\frac{r}{l^2}(r-r_1)(r^2+r_1r+r_1^2+l^2),
\end{equation}
where the second relation identifies the (only real) root $r=r_1$, namely the Schwarzschild-AdS horizon radius which satisfies
\begin{equation}
r_1^3 +l^2 (r_1 - 2m) = 0.
\end{equation}
One can check that $r_1 = r_H(\xi=0) > r_H(\xi \neq 0)$, e.g. by considering the form of $\De_r(r_H)=0$ as given in (\ref{r1rH_eqn}).  
Of most relevance here, one can show that in the small $a$ expansion
\begin{equation}
r_1-r_H 
\sim{\mathcal O}(a^2),
\end{equation}
so the regularity condition for quantities diverging like $\Delta_r^{-n}\sim(r-r_H)^{-n}$ on the horizon can be translated to $r=r_1$ at each order of the expansion, i.e. $(r-r_H)^{-1}=(r-r_1)^{-1}+(r-r_1)^{-2}{\mathcal O}(a^2)$, so we will still refer to regularity at $r=r_1$ as the ``horizon regularity condition''. The horizon angular velocity approximates to
\begin{equation}\label{Omega_H_a}
\Omega_H=\frac{a}{r_H^2}+{\mathcal O}(a^3)=\frac{a}{r_1^2}+{\mathcal O}(a^3).
\end{equation}

We will also find it useful to treat all other quantities as dimensionless by working in $m=1$ unit, so that $\Delta_0(m=1)=0$ provides the relation
\begin{equation}\label{l_r_1}
l^2=\frac{r_1^3}{2-r_1},\quad(0\leq r_1\leq2),
\end{equation}
which allows us to eliminate $l$ in each order of the small `$a$' expansion, leaving $r_1$ as the only free metric parameter. With these conventions, $r_1$ encodes both the AdS curvature and black hole sizes: $r_1=2$ ($l\rightarrow\infty,\frac{r_1}{l}=0$) is the Kerr and small black hole limit, while $r_1\rightarrow0$ ($l\rightarrow0\sim\mathcal{O}(a),\frac{r_1}{l}\rightarrow\infty$) is the highly curved AdS and large black hole limit.

\section{The AdS analog of the Blandford-Znajek split monopole}\label{BZsolution}

We turn now to the main task of obtaining an explicit solution for a force-free magnetosphere in a Kerr-AdS background. We will work in the probe approximation, ignoring the back-reaction of the magnetosphere on the geometry.\footnote{Relaxing this condition may change the asymptotic form of the geometry, as discussed recently in 
\cite{GibbonsErgoMagBH}.} For large black holes, the energy density in the magnetosphere is still sufficient to locally screen the electric field by pair production. Thus, its natural to assume that the current contributes a subleading component to the energy-momentum tensor and the magnetosphere satisfies the force-free condition, $F^{\mu\nu}J_\nu =0$, and 
\be
 T^{\mu\nu}_{;\nu} \approx T^{\mu\nu}_{\rm (EM);\nu} = F^{\nu\mu} J_\nu = 0. \label{emc}
\ee
The force-free condition implies the vanishing of the electric field in the local rest-frame of the current, and thus ${^\star}F^{\mu\nu}F_{\mu\nu}=0$, where ${^\star}F_{\mu\nu} \equiv \frac{1}{2} \varepsilon_{\mu\nu\rh\si}F^{\rh\si}$ is the dual field strength. This condition, along with axisymmetry and stationarity, implies using the general metric \eqref{metric_lapse_shift} that $A_{\varphi,\theta}A_{t,r} = A_{t,\theta}A_{\varphi,r}$, which allows the definition of the `rotation frequency' of the field \cite{BZ},
\be
 \om(\theta,\varphi) \equiv  -\frac{A_{t,\theta}}{A_{\varphi,\theta}} = - \frac{A_{t,r}}{A_{\varphi,r}}.
\ee
This implies that the independent field quantities required are $\{A_\varphi,B^\varphi,\omega\}$, which are all implicit functions of $(r,\theta)$. Moreover, as shown by Blandford and Znajek \cite{BZ}, the function $A_\varphi={\rm constant}$ specifies poloidal field surfaces, and thus $B^\varphi = B^\varphi(A_\varphi)$ and $\om=\om(A_\varphi)$. 

To obtain an explicit solution below, following BZ we will work in the small `$a$' expansion outlined above, starting from an initial radial magnetic field in the Schwarzschild limit. The physical situation assumes that the magnetic field is produced by currents in an accretion disk. A split monopole field is a crude approximation to this with opposite charge in the north and south hemispheres, allowing for a discontinuity on the equator, associated with the accretion disk. For simplicity, in the discussion below, we will not explicitly split the monopole across the equator.

\subsection{General form of the equations in the 3+1 formalism}

Rather than solve the force-free equations directly, following \cite{McKGam}, we will consider the conservation equations
 $T^{\mu\nu}_{\rm (EM);\nu} =0$ using the general metric \eqref{metric_lapse_shift}. 
Defining the shorthand notation,
\begin{align}
dT^\mu&\equiv T^{\mu\nu}_{\phantom{\mu\nu};\nu},\\
\{X,Y\}&\equiv X_{,r}Y_{,\theta}-X_{,\theta}Y_{,r}\label{brk},\\
B_T&\equiv(g_{\varphi\varphi}g_{tt}-g_{\varphi t}^2)B^\varphi\overset{BL}{=\joinrel=}-h_{\varphi\varphi}\alpha^2B^\varphi,
\end{align}
the definition of $\omega$ implies
\begin{equation}\label{brk_omega}
\{A_\varphi,\omega\}\equiv 0.
\end{equation}
We work in BL coordinates (with KS results given in Appendix \ref{Appdx_KS_results}). For $T_t^\mu$ \& $T_\varphi^\mu$ (the energy and angular momentum flux densities) we have
\begin{align}
T_t^r&=-\omega B_T\frac{A_{\varphi,\theta}}{\sqrt{-g}},\quad T_t^\theta=\omega B_T\frac{A_{\varphi,r}}{\sqrt{-g}}\label{T_tr},\\
T_t^\varphi&=\omega\bigl[\frac{\beta^\varphi(\omega+\beta^\varphi)}{\alpha^2}-\frac{1}{h_{\varphi\varphi}}\bigr]h^{MN}A_{\varphi,M}A_{\varphi,N},\\
2T_t^t&=\bigl[\frac{(\beta^\varphi)^2-\omega^2}{\alpha^2}-\frac{1}{h_{\varphi\varphi}}\bigr]h^{MN}A_{\varphi,M}A_{\varphi,N}-\frac{B_T^2}{h_{\varphi\varphi}\alpha^2},
\end{align}
and
\begin{align}
T_\varphi^r&=-T_t^r/\omega,\quad T_\varphi^\theta=-T_t^\theta/\omega\label{T_phr},\\
T_\varphi^\varphi&=-T_t^t-\frac{B_T^2}{h_{\varphi\varphi}\alpha^2},\\
T_\varphi^t&=\frac{\beta^\varphi+\omega}{\alpha^2}h^{MN}A_{\varphi,M}A_{\varphi,N}.
\end{align}
For $dT_\mu$ we have
\begin{gather}
dT_r=\frac{A_{\varphi,r}}{\sqrt{-g}}\bigl(c_\omega\sqrt{-g}h^{MN}A_{\varphi,N}\bigr)_{,M}+\frac{(h^{MN}A_{\varphi,M}A_{\varphi,N})h_{\varphi\varphi}(\beta^\varphi+\omega)\omega_{,r}+(B_T^2)_{,r}/2}{h_{\varphi\varphi}\alpha^2}\label{dT_r},\\
dT_\theta=\frac{A_{\varphi,\theta}}{\sqrt{-g}}\bigl(c_\omega\sqrt{-g}h^{MN}A_{\varphi,N}\bigr)_{,M}+\frac{(h^{MN}A_{\varphi,M}A_{\varphi,N})h_{\varphi\varphi}(\beta^\varphi+\omega)\omega_{,\theta}+(B_T^2)_{,\theta}/2}{h_{\varphi\varphi}\alpha^2}\label{dT_th},\\
dT_\varphi=-\{A_\varphi,B_T\}/\sqrt{-g}\label{dT_ph},\\
dT_t=\{A_\varphi,B_T\omega\}/\sqrt{-g}\label{dT_t},
\end{gather}
together with the following relations
\begin{align}
dT_t+\omega dT_\varphi&=B_T\{A_\varphi,\omega\}/\sqrt{-g}\overset{\eqref{brk_omega}}{=\joinrel=}0\label{dT_t+dT_ph},\\
B^idT_i&=-h^{MN}A_{\varphi,M}A_{\varphi,N}\{A_\varphi,\omega\}(\beta^\varphi+\omega)/(\sqrt{-g}\alpha^2)\overset{\eqref{brk_omega}}{=\joinrel=}0\label{B.dT},
\end{align}
where the indices $M,N=r,\theta$; $i,j=r,\theta,\varphi$ and 
\begin{equation}\label{c_omega}
c_\omega=\frac{1}{h_{\varphi\varphi}}-\frac{(\beta^\varphi+\omega)^2}{\alpha^2}.
\end{equation}
We can write the second order derivative terms in $dT_r$ \& $dT_\theta$ more concisely using a 4-D d'Alembertian:
\begin{equation}
\bar\Box A_\varphi=\frac{1}{\sqrt{-\bar g}}(\sqrt{-\bar g}\bar g^{\mu\nu}A_{\varphi,\nu})_{,\mu}=\frac{1}{c_\omega^2\sqrt{-g}}(c_\omega\sqrt{-g}g^{\mu\nu}A_{\varphi,\nu})_{,\mu}=\frac{1}{c_\omega^2\sqrt{-g}}\bigl(c_\omega\sqrt{-g}h^{MN}A_{\varphi,N}\bigr)_{,M},
\end{equation}
owing to the conditions $\partial_t(\ldots)=\partial_\varphi(\ldots)=0$, where $\bar\Box$ is associated with the metric $\bar g_{\mu\nu}$ obtained by Weyl transforming $g_{\mu\nu}$,
\begin{equation}
\bar g_{\mu\nu}=c_\omega g_{\mu\nu},\quad \bar g^{\mu\nu}=c_\omega^{-1}g^{\mu\nu},\quad \sqrt{-\bar g}=c_\omega^2\sqrt{-g}.
\end{equation}

The equations \eqref{dT_t+dT_ph} \& \eqref{B.dT} are components of the identity,
\begin{equation}
{^\star}F^{\mu\nu}dT_\nu(={^\star}F^{\mu\nu}F_{\rho\nu}J^\rho=\frac{1}{4}\,{^\star}F^{\alpha\beta}F_{\alpha\beta}J^\mu)=0,
\end{equation}
and constrain the number of independent equations in $dT_\mu=0$ from four to two. The fact that a single condition \eqref{brk_omega} gives two constraints follows from the following general argument. Namely, the existence of non-trivial solutions (i.e., $J^\mu\neq0$) to $F_{\nu\mu}J^\nu(=dT_\mu)=0$ implies $\det F_{\mu\nu}=0$ which is equivalent to the degeneracy condition (and to \eqref{brk_omega}) by the identity $\det F_{\mu\nu}=(F_{\mu\nu}{^\star}F^{\mu\nu})^2/16$ and thus the matrix $F_{\mu\nu}$ cannot have full rank 4. $F_{\mu\nu}$ thus has rank 2 since antisymmetric matrices can only have an even rank. We choose one of the independent equations to be $dT_\varphi=0$ or $dT_t=0$ which just gives the condition
\begin{equation}\label{brk_BT}
-\sqrt{-g}dT_\varphi=\{A_\varphi,B_T\}=0.
\end{equation}
Then \eqref{B.dT} determines the remaining equation to be either $dT_r=\eqref{dT_r}=0$ or $dT_\theta=\eqref{dT_th}=0$ which we focus on from now on.\footnote{Eqs.~\eqref{dT_r} and \eqref{dT_th} should correspond to the \emph{force-free Grad-Shafranov equation} as presented in, e.g., \cite{GR_GS_eqn}, if we interchange $\Psi,\widetilde\omega^2,\delta\Omega,I$ used there with  $A_\varphi,g_{\varphi\varphi},\beta^\varphi+\omega,B_T$ respectively.} In the next subsection, we will make use of the slow rotation expansion to obtain a solution. For completeness, in Appendix~\ref{AppC} we also reformulate the force-free equations in NP variables.

\subsection{Solving equations in the small `$a$' expansion}

Starting from a simple monopole solution in the Schwarzschild-AdS limit, we employ the following ansatz expanding field quantities about $a=0$ (or more precisely an expansion in $a/m$), keeping terms up to ${\mathcal{O}}(a^2)$,
\begin{align}
A_\varphi&=-Cu+a^2A_\varphi^{(2)}\label{ansatz_A_phi},\\
\omega&=a\omega^{(1)}\label{ansatz_omega},\\
B_T&=aB_T^{(1)}\label{ansatz_BT},\quad(B^\varphi=aB^\varphi_{(1)}),
\end{align}
where $C$ is proportional to the magnetic charge (if we don't `split' the monopole), and $u=\cos\theta$. Applying this ansatz to the conditions $\{A_\varphi,\omega\}=0$ and $\{A_\varphi,B_T\}=0$ yields
\begin{align}
aC\omega^{(1)}_{,r}+a^3\{A_\varphi^{(2)},\omega^{(1)}\}&=0,\\
(\omega^{(1)}\leftrightarrow B_T^{(1)})&=0.
\end{align}
Consistently dropping the ${\mathcal{O}}(a^3)$ terms\footnote{Although $a$ appears as an overall factor here, the subleading terms do indeed contribute at ${\cal O}(a^3)$ to $dT_\varphi$ and at ${\mathcal{O}}(a^4)$ to $dT_t$ and should be dropped.}, we arrive at the constraints
\begin{equation}\label{omega_BT_r-indep}
\omega_{,r}={B_T}_{,r}=0.
\end{equation}
Counting powers of `$a$' in $dT_r$ and $dT_u$ we find
\begin{align}
dT_r&=\underbrace{A_{\varphi,r}}_{{\mathcal{O}}(a^2)}c_\omega^2\underbrace{\bar\Box A_\varphi}_{{\mathcal{O}}(a^2)}+{\mathcal{O}}(1)\Bigl[\underbrace{(\beta^\varphi+\omega)\overbrace{\omega_{,r}}^{=0}}_{{\mathcal{O}}(a^2)}+\underbrace{\overbrace{(B_T^2)_{,r}}^{=0}/2}_{{\mathcal{O}}(a^2)}\Bigr]\quad\sim\quad{\mathcal{O}}(a^4),\\
dT_u&=\underbrace{A_{\varphi,u}}_{{\mathcal{O}}(1)}c_\omega^2\underbrace{\bar\Box A_\varphi}_{{\mathcal{O}}(a^2)}+{\mathcal{O}}(1)\Bigl[\underbrace{(\beta^\varphi+\omega)\omega_{,u}+(B_T^2)_{,u}/2}_{{\mathcal{O}}(a^2)}\Bigr]\quad\sim{\quad\mathcal{O}}(a^2),
\end{align}
(where it is important to note that $\bar\Box A_\varphi$ has a vanishing ${\mathcal{O}}(1)$ term which is specific to the monopole field.) Thus $dT_r=0$ is automatically solved to the desired order. The non-trivial equation $dT_u=0$ is a second order PDE for $A_\varphi^{(2)}$ and reads explicitly in the Kerr-AdS metric,\footnote{It is interesting to note that Eq.~\eqref{dT_u} does not involve $r$-derivatives of $\omega^{(1)}$ and $B_T^{(1)}$ even if we do not impose $\omega_{,r}^{(1)}=B_{T_{,r}}^{(1)}=0$. In addition, the inhomogeneous part is a total $u$-derivative, which is not obvious from the original form \eqref{dT_th} of the equation.}
\begin{multline}\label{dT_u}
dT_u=A_{\varphi,rr}^{(2)}+\frac{(1-u^2)}{\Delta_0}A_{\varphi,uu}^{(2)}+2\frac{r^3+ml^2}{l^2\Delta_0}A_{\varphi,r}^{(2)}\\
+\frac{1}{2C\Delta_0^2}\Bigl[C^2(1-u^2)^2\bigl[r^4(\omega^{(1)})^2+\bigl(\frac{r^2}{l^2}-\frac{2m}{r}\bigr)(2r^2\omega^{(1)}-1)\bigr]-r^4(B_T^{(1)})^2\Bigr]_{,u},
\end{multline}
where $\Delta_0$ is given in \eqref{Delta0}.

To solve \eqref{dT_u}, we employ a separation of variables $A_\varphi^{(2)}=f(r)g(u)$. For terms in the inhomogeneous part with different powers of $r$ to have the same $u$-dependence, namely $u(1-u^2)$, and for $\omega^{(1)}(u)$ \& $B^\varphi_{(1)}(u)\sim B_T^{(1)}(u)/(1-u^2)$ to be regular at $u=1$, we require
\begin{align}
\omega^{(1)}(u)&=\omega^{(1)},\\
B_T^{(1)}(u)&=B_T^c(1-u^2),
\end{align}
where $\omega^{(1)}$ and $B_T^c$ are constants. Meanwhile, we can fix
\begin{equation}
g(u)=u(1-u^2).
\end{equation}
The remaining radial equation is
\begin{multline}\label{fr_eqn0}
f''(r)+\frac{2r^3+r_1(r_1^2+l^2)}{r(r-r_1)(r^2+r_1r+r_1^2+l^2)}f'(r)-\frac{6l^2f(r)}{r(r-r_1)(r^2+r_1r+r_1^2+l^2)}\\
+2Cl^2\frac{\{[C^{-2}(B_T^c)^2-(\omega^{(1)})^2]l^2-2\omega^{(1)}\}r^5+r^3+(2\omega^{(1)}r^2-1)r_1(r_1^2+l^2)}{r^3(r-r_1)^2\bigl(r^2+r_1r+r_1^2+l^2\bigr)^2}=0.
\end{multline}
At this point, we can eliminate one of the $(r-r_1)^{-2}$ factors in the inhomogeneous part by choosing
\begin{equation}\label{BTsq_omega}
(B_T^c)^2=C^2\Bigl(\omega^{(1)}-\frac{1}{r_1^2}\Bigr)^2.
\end{equation}
Though this may appear ad hoc, it is actually equivalent to BZ's horizon regularity condition \cite{BZ}. The same relation can be obtained more rigorously from the regularity of $\tilde B^\varphi$ in KS coordinates, as presented in Appendix \ref{Appdx_KS_results}, where the sign ambiguity of $B_T^c$ in \eqref{BTsq_omega} is also fixed, given by \eqref{BT_omega_A_phi_a} \& \eqref{BT_omega}:
\begin{equation}\label{BT1_omega1}
B_T^c=-\bigl(\omega^{(1)}-\frac{\Omega_H}{a}\bigr)A_{\varphi,u}+\mathcal{O}(a^3)=C\Bigl(\omega^{(1)}-\frac{1}{r_1^2}\Bigr)+\mathcal{O}(a^3).
\end{equation}

We proceed with $B_T^c$ fixed via \eqref{BTsq_omega}. Transforming to a dimensionless radial coordinate
\begin{equation}\label{z_coord}
z\equiv r_1/r,
\end{equation}
and using $m=1$ units to eliminate $l$, we arrive at
\begin{multline}\label{fz_eqn}
f''(z)+\frac{2z(3z-r_1)}{(z-1)[2z^2-(r_1-2)(z+1)]}f'(z)+\frac{6r_1f(z)}{(z-1)[2z^2-(r_1-2)(z+1)]}\\
-C\frac{4(z^2-2\omega^{(1)}r_1^2)(z^2+z+1)+2r_1(z+1)}{(z-1)[2z^2-(r_1-2)(z+1)]^2r_1}=0.
\end{multline}
The horizon is at $z=1$ and spatial infinity at $z=0$.

A comparison with the Kerr case is in order. In the Kerr limit ($r_1=2$), equation \eqref{fz_eqn} develops a second singular point $z=0$ (besides $z=1$) near which it behaves like
\begin{multline}
f''(z)+[2z^{-1}+{\mathcal{O}}(1)]f'(z)-[6z^{-2}+{\mathcal{O}}(z^{-1})]f(z)\\
-\frac{C}{2}z^{-4}(8\omega^{(1)}-1)(1+2z+3z^2)+{\mathcal{O}}(z^{-1})=0.
\end{multline}
The leading ${\mathcal{O}}(z^{-4})$ divergence of the inhomogeneous part can only be removed by choosing $\omega^{(1)}=1/8$, a fixed rotation frequency which is half the horizon angular velocity. For the general Kerr-AdS case ($r_1\ne 2$), the equation is well-behaved at $z=0$ and no obvious constraint on $\omega^{(1)}$ is needed. This is the first, and perhaps most significant, difference we observe in the properties of the force-free magnetosphere in the Kerr-AdS background.

\subsection{Series and numerical solutions}

For the Kerr geometry, the indicial equation implies integer asymptotics $f(z) \sim z^0, z^1$ and due to the singular point there is logarithmic scaling. Requiring the boundary condition $f(0)=0$, the expansion for the homogeneous equation has the form
\begin{equation}
f(z)_\text{Kerr}=\sum_{n=1}^{\infty}\bigl[c_n+c_n'\ln(z)\bigr]z^n,\quad(c_1'=0).
\end{equation}
It turns out that the inhomogeneous term is only consistent with this regular scaling at infinity with the unique choice of $\omega^{(1)}=1/(8m^2)$ noted above. An analytic solution for $f(z)$ can then be obtained in terms of dilogarithms. Blandford and Znajek used matching condition at infinity that we will return to later to obtain the same result \cite{BZ}.

\begin{figure}[tbp]
\centering\includegraphics[scale=0.4]{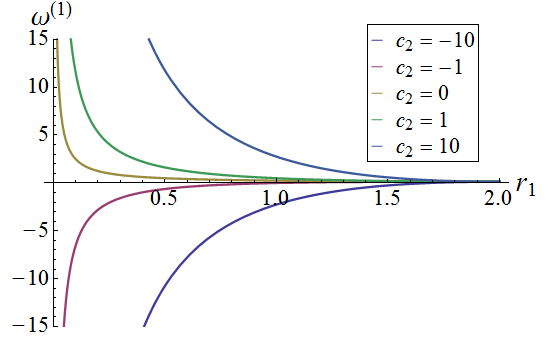}\includegraphics[scale=0.4]{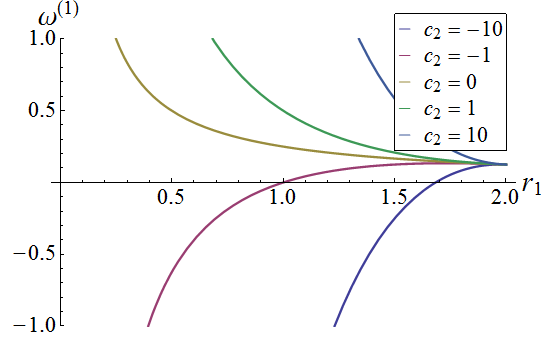}
\caption{Plots of $\omega^{(1)}$ as functions of $r_1$ for various values of $c_2$. The right-hand panel shows the region close to the $r_1$-axis.}
\label{omega}
\end{figure}

For the general Kerr-AdS case, although an analytic solution does not appear possible - barring a perturbative expansion about the Kerr limit discussed in Appendix~\ref{AppB} - we can proceed in the same way since the indicial equation has the same form. The lack of an additional singular point in this case implies the existence of a regular series solution about $z=0$, and we again fix the boundary condition $f(0)=0$ corresponding to the normalizable mode in AdS, 
\begin{equation}\label{fz_series}
f_c(z)=\sum_{n=1}^{\infty}c_nz^n.
\end{equation}
In this case, the inhomogeneous term is nonsingular away from the Kerr limit, and thus we do not obtain a unique constraint on the field angular velocity $\om$. 
Substituting \eqref{fz_series} into \eqref{fz_eqn} and expanding in $z$, we obtain recursion relations expressing $\omega^{(1)}$ and $c_n$ ($n\geq3$) in terms of $c_1$ and $c_2$:
\begin{align}
\omega^{(1)}&=\frac{1}{4r_1}+\frac{c_2}{C}\frac{(r_1-2)^2}{4r_1}\label{omega1_c2_m=1},\\
c_3&=\frac{2c_1r_1}{3(2-r_1)},\\
c_n&=c_n(c_1,c_2),\quad(n>3).
\end{align}
Note $\omega^{(1)}$ depends only on $c_2$ (concavity of $f(z)$ at $z=0$) and $c_3$ only on $c_1$. We will therefore use $\omega^{(1)}$ and $c_2$ interchangeably below, and the explicit relation is shown for various parameters in Fig.~\ref{omega}.

For comparison with the numerical solution to be discussed below, it is also useful to consider a second series solution constructed about the singular point $z=1$, i.e., the horizon, where we explicitly demand the absence of the logarithmic term:
\begin{equation}\label{fz_series_1}
f_b(z)=\sum_{n=0}^{\infty}b_n(z-1)^n,
\end{equation}
with
\begin{align}
\omega^{(1)}
&=\frac{b_0}{C}\frac{r_1-3}{2}-\frac{b_1}{C}\frac{(r_1-3)^2}{6r_1}+\frac{r_1+3}{6r_1^2},\\
b_n&=b_n(b_0,b_1),\quad n\geq2.
\end{align}
The free parameters \{$\omega^{(1)},c_1$\} in $f_c(z)$ and \{$\omega^{(1)},b_0$\} in $f_b(z)$ are related through two boundary conditions $f_b(0)=0$ \& $f_b'(0)=c_1$ imposed on the series $f_b(z)$ whose radius of convergence covers $z=0$.\footnote{Among the boundary conditions $f_b(0)=0$, $f_b'(0)=c_1$, $f_c(1)=b_0$ \& $f_c'(1)=b_1$, only the first two are consistent with the value of $c_1$ obtained by BZ in the Kerr limit.} This leaves only one free parameter which we take to be $\omega^{(1)}$ (or equivalently $c_2$), with
\begin{equation}\label{c1b0_r1c2}
c_1(r_1,c_2)\text{ or }b_0(r_1,c_2)=p(r_1)-q(r_1)c_2
\end{equation}
where $p(r_1)$ \& $q(r_1)$ are ratios of polynomials in $r_1$. Matching the expansion e.g. at $z=0$ reproduces the numerical results discussed below to high precision.

\begin{figure}[tbp]
\centering\includegraphics[scale=0.5]{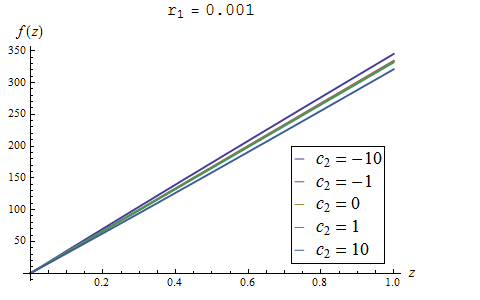}\includegraphics[scale=0.5]{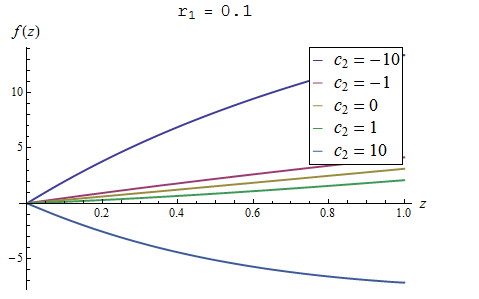}\\
\includegraphics[scale=0.5]{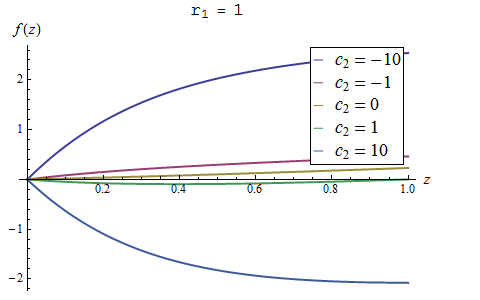}\includegraphics[scale=0.5]{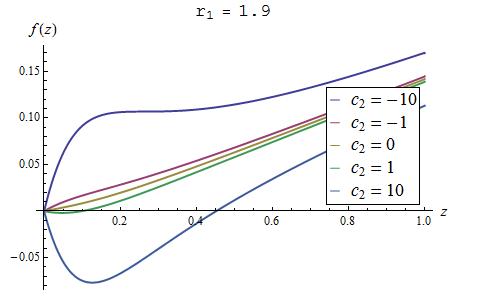}\\
\includegraphics[scale=0.5]{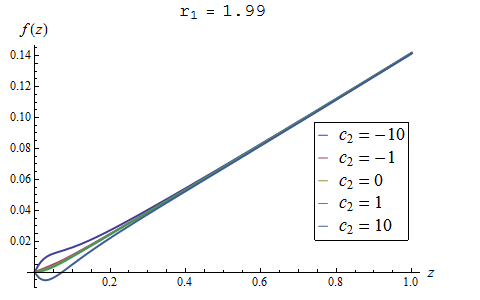}\includegraphics[scale=0.5]{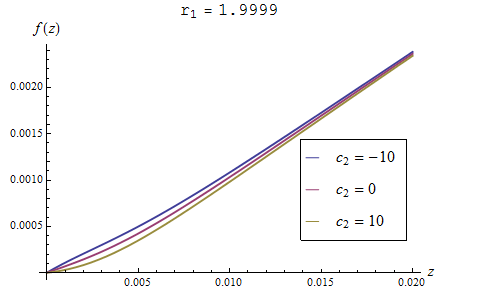}
\caption{In each graph (with fixed $r_1$), a set of solution curves corresponding to various choices of $\omega^{(1)}=\frac{1}{4r_1}+\frac{c_2}{C}\frac{(r_1-2)^2}{4r_1}$ are shown, by varying $c_2$ from $-10$ to $10$ (setting $C=1$). Note that for a large black hole with $r_1 = 0.001$, $\omega^{(1)}$ varies by a relative factor of 100 as $c_2$ is varied, while for the small black hole with $r_1=1.9999$, $\om^{(1)}$ only varies by a relative factor of $10^{-7}$. 
Thus the curves effectively zoom in to the ``middle curve'' with $c_2=0$ as $r_1$ increases, and $c_2=0$ is indeed the only solution in the Kerr limit. The ``middle curve'' is plotted for various $r_1$'s in Fig.~\ref{omega1=1_4r1}.}\label{curves_c2_r1}
\end{figure}

\begin{figure}[tbp]
\centering\includegraphics[scale=0.5]{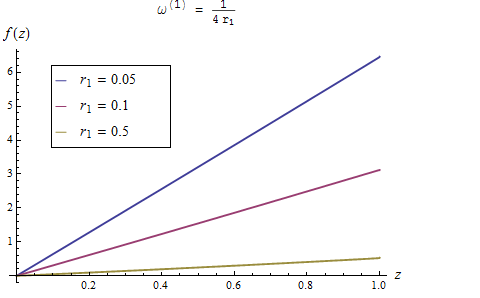}\includegraphics[scale=0.5]{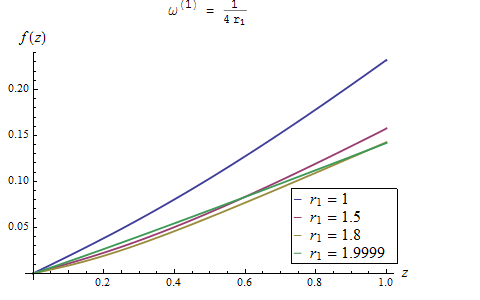}
\caption{Solution curves with $\omega^{(1)}=1/(4r_1)$ (the middle curve in each graph of Fig. \ref{curves_c2_r1}) for various $r_1$'s. Note both $c_1=f'(0)$ and $b_0=f(1)$ decrease monotonically with increasing $r_1$, while the middle part of the curve bounces back when $r_1\rightarrow2$.}\label{omega1=1_4r1}
\end{figure}

The equation for $f(z)$ can be solved numerically by shooting from the boundary to the horizon for each value of $\omega^{(1)}$ (or equivalently $c_2$), by tuning the value of $c_1$ (or $b_0$) until we get a regular solution near $z=1$. This fixes the final integration constant and, as noted above, the values of $c_1$ and $b_0$ are numerically close to those determined through direct analysis of the series solution using \eqref{c1b0_r1c2}. Plots of these solutions are shown in Fig. \ref{curves_c2_r1} for a range of different $r_1$ values, and for each $r_1$ we show a set of curves labelled by $\omega^{(1)}$ (or more conveniently $c_2$). As the plots show, for each arbitrarily picked $c_2$, a unique solution curve can be found that is regular at $z=1$ and satisfies $f(0)=0$. This agrees with the above analysis using series solutions, namely  that the boundary conditions alone do not put any constraints on $\omega$. Note that the curve with $c_2=0$, that asymptotes to the Kerr solution in the small black hole limit, has $\omega=a/(4r_1)\leq\Omega_H=a/r_1^2$ for $0<r_1\leq2$.

\subsection{Energy-momentum flux in the BZ process}\label{BZ_fluxes}

We can evaluate the relevant energy and angular momentum densities, $T^r_t=-\omega T^r_\varphi=r^{-2}\omega(\omega-\Omega_H)A_{\varphi,\theta}^2$, from Eqs.~\eqref{T_tr} \& \eqref{BT1_omega1}. Then Eqs.~\eqref{horizn_ineq} \& \eqref{1st2nd_law} imply
\begin{equation}
T\delta S\propto T^r_t+\Omega_HT^r_\varphi=r^{-2}(\omega-\Omega_H)^2A_{\varphi,\theta}^2\geq0,
\end{equation}
which always holds. Meanwhile, using Eq.~\eqref{fluxes_inequality},
\begin{align}
\delta E&\propto T^r_t+\Omega T^r_\varphi=r^{-2}(\omega-\Omega_H)(\omega-\Omega)A_{\varphi,\theta}^2,\\
\delta L&\propto-T^r_\varphi=r^{-2}(\omega-\Omega_H)A_{\varphi,\theta}^2,
\end{align}
so for energy extraction $\delta E<0$ we require either
\begin{equation}\label{Omega<Omega_H}
\Omega<\omega<\Omega_H,\quad\delta L<0,
\end{equation}
or
\begin{equation}\label{Omega_H<Omega}
\Omega_H<\omega<\Omega,\quad\delta L>0.
\end{equation}
The energy-defining Killing vector with $\Omega=\Omega_H$ results in a 1st law without the  $\delta L$ term so that $\delta E=T\delta S\propto(\omega-\Omega_H)^2\geq0$, and thus no energy extraction, regardless of the value of $\omega$. This is of course consistent with the stability arguments discussed in earlier sections. However, if we define energy using the Killing vector with $\Omega=\Omega_\infty$, we find that `energy'  can be extracted if $\omega$ satisfies the condition \eqref{Omega<Omega_H}, which is usually expected as in the Kerr case, and we have $\delta E\propto(\omega+a/l^2)(\omega-\Omega_H)<0$. If $\omega$ is outside the range \eqref{Omega<Omega_H} we have field lines rotating either backwards (even as seen in the non-rotating frame at infinity) or faster than the black hole, and there is no energy extraction.

In conclusion, the stability condition apparently implies that any choice of energy-defining Killing vector, other than via $\Om_H$, leads to a rather benign form of `ergosphere' and BZ process.  An outgoing radial Poynting flux is possible, but does not reflect an instability or spin-down of the black hole, as it can be turned off by switching to an alternate definition of energy for an asymptotic observer. In effect, there is still a net ingoing flux when one properly accounts for both energy and angular momentum. Nonetheless, the fact that the AdS/CFT correspondence points to a specific definition of `energy' which apparently exhibits this benign ergosphere in the bulk raises the question of how it is reflected (if at all) in the dual field theory. We will turn to this question shortly, after first considering in more detail whether there are additional constraints on the field rotation velocity $\om$.

\section{Asymptotic matching to a rotating monopole in AdS space-time}\label{matching}

The class of solutions obtained above was parametrized by the field angular velocity $\om$. This contrasts with the Kerr limit, where a unique choice of $\om$ is required for regularity. In this section, we consider whether there are analogous constraints imposed by demanding that the solution match asymptotically at large radius onto a rotating monopole in AdS. We will proceed in this section to find an exact rotating monopole solution in AdS, an analog of the Michel solution in flat space \cite{Michel}. In practice, since the AdS boundary is only defined up to a Weyl scaling, the definition of the asymptotic monopole is ambiguous due to possible $\mathcal{O}(a^2)$ corrections associated with squashing consistent with axisymmetry. In addition, there is also the possibility to add further $r$-dependent $\mathcal{O}(a^2)$ corrections, that change the value of $\om$ obtained by matching. Thus, ultimately we find that the full range of $\om$ obtained in the Kerr-AdS solutions above can still be consistently matched to an asymptotic `monopole' in AdS.

\subsection{Matching to a  perturbed monopole}
One can show that the arbitrariness in the value of $\omega$ persists for solutions in pure AdS space. This again contrasts with flat spacetime where the unique solution is Michel's rotating monopole \cite{Michel}. The freedom to choose $\om$ in the Kerr-AdS case is reflected in the asymptotic $dT_u=0$ equation by the presence of two contributions: the inhomogeneous part ($\sim C$, from the monopole $-Cu$ alone) and $c_2=f''(z=0)$ from the $\mathcal{O}(a^2)$ correction. We thus consider a rotating monopole ansatz in AdS space (denoted with a bar) allowing for possible $\mathcal{O}(a^2)$ corrections,
\begin{equation}\label{A_Phi_pertb}
\bar A_\Phi(y,U)=-CU+a^2U(1-U^2)\bar f(y),\quad\bar B_T(U)=a(1-U^2)\bar B_T^c, 
\end{equation}
using coordinates $\{y,U=\cos\Theta,\Phi,T\}$ in which the AdS metric assumes the standard form
\begin{equation}\label{AdS_U}
{\operatorname{d}s}^2=\bigl(1+\frac{y^2}{l^2}\bigr)^{-1}{\operatorname{d}y}^2+\frac{y^2}{1-U^2}{\operatorname{d}U}^2+y^2(1-U^2){\operatorname{d}\Phi}^2-\bigl(1+\frac{y^2}{l^2}\bigr){\operatorname{d}T}^2.
\end{equation}
The equation for $\bar f(y)$ for small `$a$' reads,\footnote{It is worth noting that the equation takes a simpler form $f''(x)-6f(x)/\sin^2x+2Cl^2[(\bar B_T^c/C)^2-\bar\omega^2]=0$ if we define the new radial coordinate $x\equiv\arctan(y/l)$.}
\begin{equation}
f''(y)+\frac{2y}{y^2+l^2}f'(y)-\frac{6l^2}{y^2(y^2+l^2)}f(y)+2Cl^4\frac{(\bar B_T^c/C)^2-\bar\omega^2}{(y^2+l^2)^2}=0,
\end{equation}
which has an analytic solution
\begin{equation}
f(y)=-\frac{\bar c_2}{4\pi y^2}\Bigl[2\pi\bigl(\frac{y^2}{l^2}+3\bigr)\arctan^2\frac{y}{l}-\bigl[(\pi^2+12)\bigl(\frac{y^2}{l^2}+3\bigr)+12\pi\frac{y}{l}\bigr]\arctan\frac{y}{l}+3\frac{y}{l}(2\pi\frac{y}{l}+\pi^2+12)\Bigr],
\end{equation}
where
\begin{equation}\label{bar_c2}
\bar c_2=-Cl^4\bigl[(\bar B_T^c/C)^2-\bar\omega^2\bigr]
\end{equation}
is an overall free parameter which turns out to be the coefficient of the $\mathcal{O}(y^{-2})$ term in the expansion,
\begin{equation}
f(y)=\bar c_2\Bigl[\frac{\pi^2-12}{\pi l}y^{-1}+y^{-2}+\mathcal{O}(y^{-3})\Bigr].
\end{equation}
The $\mathcal{O}(a^2)$ perturbation is only significant around $y\sim l$.
The ansatz \eqref{A_Phi_pertb} without the $\mathcal{O}(a^2)$ correction yields an exact solution (for any $a$) analogous to Michel's solution, with the relation $B_T^c=-C\bar\omega^{(1)}$ (further imposing $\bar B_T(U=1)=0$ and dropping the possible $U$-dependence of $\bar\omega^{(1)}$ for it to match the constant value in BZ solution).

We try matching the full Kerr-AdS solution to this perturbed rotating monopole at large radius. First note that \eqref{AdS_U} is related to the large-$r$ (or zero-mass) Kerr-AdS metric by coordinate transformations \cite{Hawking_rotn_AdSCFT} which take the form
\begin{equation}\label{zero-mass_AdS_r}
y=r\sqrt{\frac{\Delta_\theta}{\Xi}},\quad U=u\sqrt{\frac{\Xi}{\Delta_\theta}},\quad\Phi=\varphi+\frac{a}{l^2}t,\quad T=t,
\end{equation}
at large $r$. Using \eqref{zero-mass_AdS_r}, the matching requires that we equate
\begin{equation}\label{match_BT_omega}
\bar B_T^c=B_T^c \text{ (which holds in small `$a$' limit)},\quad \bar\omega^{(1)}=\omega^{(1)}+\frac{1}{l^2},
\end{equation}
and we obtain
\begin{equation}\label{omega1_c2bar}
\omega^{(1)}=\frac{l^2-r_1^2}{2l^2r_1^2}+\frac{\bar c_2}{C}\frac{r_1^2}{2l^2(l^2+r_1^2)}\overset{m=1}{=\joinrel=\joinrel=}\frac{r_1-1}{r_1^3}+\frac{\bar c_2}{C}\frac{(r_1-2)^2}{4r_1^3},
\end{equation}
which is analogous to \eqref{omega1_c2_m=1}. Thus even after the matching we still have one free parameter $\bar c_2$ or $c_2$ that renders $\omega$ arbitrary.

Naively the rotating monopole in AdS is able to produce energy and momentum fluxes: $T_T^y=-\bar\omega T_\Phi^y=C\bar\omega\bar B_T/y^2$. This is not surprising if we expect it to serve as the asymptotic limit of the interior BZ process. Note that the fluxes are singular at $y=0$, but this solution needs to be interpreted with a physical cut-off such as the surface of a star.

For completeness, we also present the currents (assuming the same matching conditions):
\begin{align}
J^y&=-\frac{aU}{l^2y^2}\Bigl(C\frac{r_1^2+l^2}{r_1^2}-\bar c_2\frac{r_1^2}{r_1^2+l^2}\Bigr)\overset{m=1}{=\joinrel=\joinrel=}-\frac{aU}{2r_1^3y^2}\bigl[4C-\bar c_2(r_1-2)^2\bigr],\\
J^\Phi&=-2a^2U\frac{\bar c_2}{l^2y^2(l^2+y^2)}\overset{m=1}{=\joinrel=\joinrel=}-2a^2U\frac{\bar c_2(r_1-2)^2}{r_1^3y^2[y^2(2-r_1)+r_1^3]},\qquad J^U=0,\\
J^T&=-\frac{aU}{y^2(l^2+y^2)}\Bigl(C\frac{r_1^2+l^2}{r_1^2}+\bar c_2\frac{r_1^2}{r_1^2+l^2}\Bigr)\overset{m=1}{=\joinrel=\joinrel=}-\frac{aU}{2y^2[y^2(2-r_1)+r_1^3]}\bigl[4C+\bar c_2(r_1-2)^2\bigr].
\end{align}
We have separated contributions from the monopole ($\sim C$) and the $\mathcal{O}(a^2)$ correction ($\sim\bar c_2$). Note that $J^\Phi$ only contains $\bar c_2$.

As another example, one can also consider the gauge potential $A_\Phi$ in the exact KNAdS `vacuum' solution, expanded for small `$a$', $A_\Phi=-Cu[1+a^2(1/l^2+(1-u^2)/r^2)]=-CU\{1+a^2[(3-U^2)/(2l^2)+(1-U^2)/y^2]\}$. One obtains a similar configuration that has energy and momentum fluxes.

\subsection{Matching with an unperturbed monopole}

\begin{figure}[tbp]
\centering\includegraphics[scale=0.6]{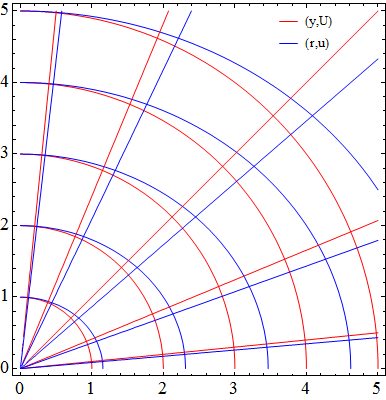}
\caption{Poloidal coordinates for AdS are shown, indicating the asymptotic squashing of the 2-sphere in BL coordinates, with the vertical axis being the rotation axis. Red curves/lines are constant $(y,U)$ grids and blue curves/lines are constant $(r,u)$ grids using the same set of constants, e.g., $U=u=\cos\frac{\pi}{4}$ are shown together. The blue (BL) grid deforms away from those of a unit 2-sphere (times the radial direction).}\label{yUru_coords}
\end{figure}

We can simplify the matching by considering only the unperturbed monopole in AdS. However a subtlety not present in the Kerr case is that, due to the non-trivial $U\leftrightarrow u$ transformation, what we called a monopole (i.e., $-Cu$) in our perturbative ansatz is not quite the same object as $\bar A_\Phi(U)=-CU$. When there is no rotation the two `monopoles' are the same ($U=u$ for $a=0$). When the black hole is spun up, an observer at infinity may adopt one of the following two reference frames:
\begin{enumerate}
\item The asymptotic observer sees a rotating field given by the exact solution $-CU$ as if the monopole is spun up in a fixed pure AdS background with the standard metric in $(y,U)$ coordinates (apart from a constant shift in $\Phi$ by frame-dragging). This observer does not know that an interior observer would have switched to $(r,u)$ coordinates by insisting on the horizon being a constant-$r$ surface (in BL coordinates).
\item The asymptotic observer does account for the change in their local geometry caused by the black hole rotation and agrees with the interior observer who describes the monopole as $-Cu$. The observer at infinity would then use new coordinates $(y',U')$ to recognize the standard AdS geometry of the boundary, in accordance with the change in the shape of the horizon in the bulk.
\end{enumerate}
The coordinate grids of $(y,U)$ \& $(r,u)$ systems are sketched in Fig.~\ref{yUru_coords} showing their relative deformation so that the monopole naturally defined in one system will appear to have non-uniformly distributed radial field lines as seen in another.

To match in the first case, one needs to replace $-Cu$ with $-CU=-Cu+Cu(1-u^2)\frac{a^2}{2l^2}+\mathcal{O}(a^4)$ for the interior ansatz. This does not affect the relation \eqref{BT1_omega1} between $B_T^c$ \& $\omega^{(1)}$ obtained from the horizon regularity condition but modifies \eqref{omega1_c2_m=1} slightly (replacing $f(r)\rightarrow f(r)+\frac{C}{2l^2}$). Setting $\bar c_2=0$ in \eqref{omega1_c2bar} we find
\begin{equation}
\omega^{(1)}=\frac{1}{2r_1^2}-\frac{1}{2l^2}\overset{m=1}{=\joinrel=\joinrel=}\frac{r_1-1}{r_1^3},\qquad c_2=C/(2r_1^2).
\end{equation}
Note that $\omega-\Omega_\infty=\frac{1}{2}(\Omega_H-\Omega_\infty)>0$, which is analogous to $\omega=\Omega_H/2$ in the Kerr case.

To match in the second case, we look for $\bar B_T$ \& $\bar\omega$ corresponding to $-Cu$ directly in the large-$r$ (or zero-mass) Kerr-AdS metric, but this is just equivalent to setting $A_\varphi^{(2)}\sim f(r)=0$ in our perturbative ansatz and solving for the large-$r$ (small-$z$) limit of \eqref{fz_eqn}. We obtain
\begin{equation}
\omega^{(1)}=\frac{l^2}{2r_1^2(r_1^2+l^2)}=\frac{1}{4r_1m},
\end{equation}
corresponding to \eqref{omega1_c2_m=1} with $c_2=0$. In both cases $\Omega_\infty<\omega<\Omega_H$ satisfying the condition for energy extraction.

\section{Aspects of the Dual Field Theory}\label{dualfield}

The near-equilibrium behaviour of `large' AdS black holes, satisfying $r_H > l$, has a dual holographic description in terms of the grand canonical ensemble for a field theory on the 2+1-dimensional boundary geometry. For rotating black holes, this system is characterized by a fluid at finite temperature on a rotating two-sphere, and the force-free magnetosphere we have studied translates to an electromagnetic perturbation of this rotating fluid. We will focus attention on large black holes with $r_H>l$ for the rest of this section.

To gain insight into the properties of this system, it is useful to compare with the corresponding Kerr-Newman-AdS (KNAdS) vacuum solution of Einstein-Maxwell theory. In the slow-rotation regime, and using BL coordinates, the KNAdS geometry with magnetic charge $C$ admits the expansion
\begin{align}
 {\rm d}s^2_{\rm KNAdS}& = {\rm d}s^2_{\rm KAdS} + {\cal O}(C^2) \\
 A_{\rm KNAdS}& = -C \cos\theta {\rm d}\varphi + {\cal O}(a).
\end{align}
Thus, by working to linear order in $C$ (and thus ignoring magnetic back-reaction on the Kerr-AdS geometry), we can equivalently consider the force-free solution as a particular ${\cal O}(a/m)$ perturbation of the KNAdS background. The first difference emerges in the ${\cal O}(a)$ correction to $A_t$. For KNAdS, $A_t = Ca\cos\theta/r^2 +{\cal O}(a^3)$ and this completes the solution up to ${\cal O}(a^3)$. In contrast, the force-free solution necessarily satisfies $A_t = -\omega A_\varphi =  C \omega^{(1)} a \cos\theta +{\cal O}(a^2)$, up to a possible constant shift  in $A_t$ that, as we will see below, can be gauged away. With $\omega$ constant, this produces a non-vanishing boundary limit for $A_t$. The holographic dictionary generally allows us to isolate the chemical potential and charge density from the asymptotics of $A_t \sim \mu + \rho/r + \cdots$ (constant coefficients will be absorbed in this identification, since the bulk gauge coupling is arbitrary in the limit that we ignore back-reaction on the metric). In the present case, this would lead to the odd conclusion that the fluid on the sphere had a $\theta$-dependent chemical potential, but no charge density to ${\cal O}(a^2)$. However, the full definition of the chemical potential \cite{mhd},
\begin{equation}
\mu l  
   = A_\mu K_{\Omega_H}^\mu |_{r\rightarrow \infty} - A_\mu K_{\Omega_H}^\mu |_{r\rightarrow r_H} = 0 + {\cal O}(a^3),
\end{equation}
does in fact vanish to this order, as expected.

This discussion suggests that the distinction between the holographic dual of the force-free magnetosphere and KNAdS may in effect be rather minor up to ${\cal O}(a)$. There is no azimuthal current at this order, and thus the boundary fluid rotates in the leading-order monopole magnetic field. The absence of a charge density at this order appears consistent with the conclusion that the angular velocity $\om$ of the electromagnetic field was not uniquely fixed, at least to ${\cal O}(a)$, in the solution. Thus there is no restriction to rotating a magnetic field in a neutral fluid. Another viewpoint follows from noting that for an electromagnetic field strength $f_{\mu\nu}$ in 2+1 dimensions, det$(f_{\mu\nu})=0$ identically, and thus the force-free condition for the dual fluid $f_{\mu\nu}j^\nu=0$ can always be solved for a specific current configuration independent of the background field. There is no analog of the constraint tr$({^\star}FF)=0$ required in 3+1 dimensions. This opens the possibility that some of the above conclusions may actually extend to higher orders in the expansion in the rotation parameter $a$, where the vanishing of the charge density need no longer hold. This also suggests that repeating the calculation in one higher dimension, where the boundary force-free condition is less trivial, may lead to somewhat different conclusions.

The arguments above imply that, at least to ${\cal O}(a)$, we can directly translate various results from the equilibrium thermodynamics of KNAdS duals to the force-free solution. In fact, to linear order in the magnetic charge, we can adopt KAdS relations for the field theory temperature,
 \begin{align}
 T 
   &\sim \frac{r_1}{4\pi l^2} \left( 1 + \frac{l^2}{r_1^2}\right) + {\cal O}(a^2)
       \;\stackrel{m=1}{\longrightarrow} \;\frac{1}{2\pi r_1^2} + {\cal O}(a^2),
\end{align}
and the angular velocity
\begin{align}
 \Om = \Om_H - \Om_\infty 
  &\sim \frac{a}{l^2}\left(1 + \frac{l^2}{r_1^2}\right) + {\cal O}(a^3)
     \; \stackrel{m=1}{\longrightarrow} \; \frac{2a}{r_1^3} + {\cal O}(a^3).
\end{align}
The two quantities here $\Om_H$ and $\Om_\infty$ are the bulk angular velocities at $r=r_H$ and $r=\infty$ respectively. The dual field theory is identified as a neutral fluid  with temperature $T$ in a rotating Einstein universe, with angular velocity $\Om$. These quantities along with suitable definitions of mass and angular momentum then satisfy the first law, as discussed earlier \cite{Gibbons1stlaw}.  Indeed the partition function of an ideal gas in this background can be computed and reproduces the structure of the bulk partition function \cite{Hawking_rotn_AdSCFT,Thermo_KerrAdS_CFT,LargeAdSfluid}, which for KNAdS has the form \cite{mhd},
\be
 \frac{1}{V} \ln Z = T^2 \frac{h\left(\mu/T,B(1-\Om^2l^2)/T^2\right)}{1-\Om^2 l^2} \sim T^2 h\left(\mu/T,B/T^2\right) + {\cal O}(a^2).
\ee
The quantity $h(\mu/T,B/T^2)$ specifies the partition function of the static charged black hole. Note that the free energy diverges, and the rotation velocity exceeds the speed of light, unless $\Om < 1/l$ \cite{Hawking_rotn_AdSCFT}. In the slow rotation limit, this condition is always satisfied for $a < r_1/2$ given that we require $r_1>l$ to have a dual description in field theory.

\subsection{Currents at $\mathcal O(a^2)$}

\begin{figure}[tbp]
\centering\includegraphics[scale=0.4]{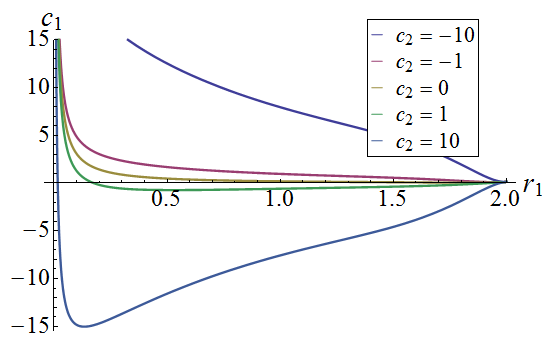}\includegraphics[scale=0.4]{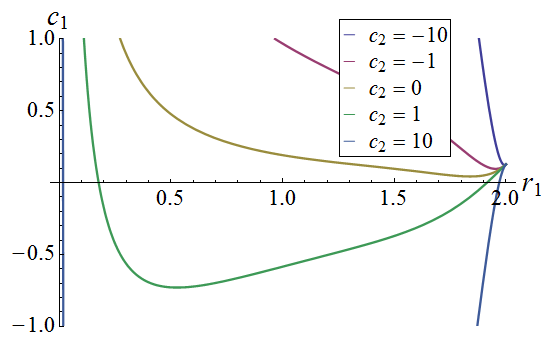}\\
\caption{Plots of the azimuthal current $c_1$ as functions of $r_1$ for various values of $c_2$, with $c_1$ obtained by solving the boundary conditions $f_b(0)=0$ and $f_b'(0)=c_1$ for the series solution $f_b(z)$ constructed up to $\mathcal{O}((z-1)^{10})$. Note that the boundary field theory interpretation only holds for large black holes with $r_1 <1$. The right-hand panel shows the region close to the $r_1$-axis. For $c_1$, good agreement is found between the numerical and series results, except for the non-monotonic behaviour of $c_1$ near $r_1\approx2$.}\label{c1b0_r1c2_plot}
\end{figure}

The corrections to the free energy arise at ${\cal O}(a^2)$, and thus are only fully calculable on accounting for the back-reaction which starts at this order. However, we can look again at the electromagnetic field, and ask whether this picture of a neutral dual fluid persists to higher orders in $a$. In fact, since $A_t = -\om A_\varphi$ is generic for axisymmetric force-free solutions, the ${\cal O}(a^2)$ correction to $A_\varphi$ characterized by $f(r)\rightarrow {\cal O}(1/r)$ has the right falloff to produce  a contribution to the charge density at ${\cal O}(a^3)$. However, we would need to compute the full solution at this order to test whether this remains or is cancelled by other terms.

Nonetheless, since the bulk solution is valid at ${\cal O}(a^2)$, we can read off the corresponding boundary currents from the asmptotics of the gauge field. In particular, with $f(z)\rightarrow c_1z$ we can (up to normalization), identify the the azimuthal boundary current,
\be
 j_\varphi= a^2 c_1, \;\;\; ({\rm given}\; r_H > l),  
 \ee
 with the results plotted in Fig. \ref{c1b0_r1c2_plot} for various choices of $r_1$ and with $\frac{c_2}{C}=-10,-1,0,1,10$. As noted above, the field theory charge density $j_t = -\om j_\varphi$ (taking the asymptotic limit of $\ptl_r A_t = -\om \ptl_r A_\varphi$) is of higher order, ${\cal O}(a^3)$, and thus we cannot perform a nontrivial test of the putative `boundary force-free condition' that is hinted at by the results at ${\cal O}(a)$.

\begin{figure}[tbp]
\centering\includegraphics[scale=0.4]{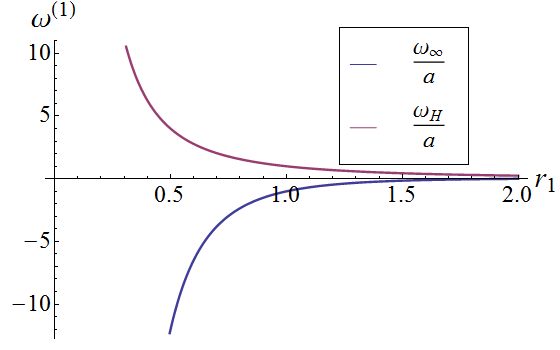}\includegraphics[scale=0.4]{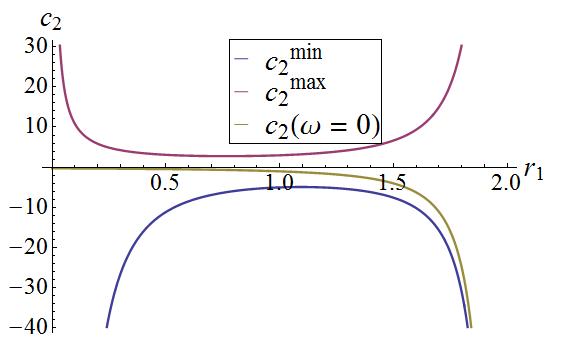}\\
\includegraphics[scale=0.4]{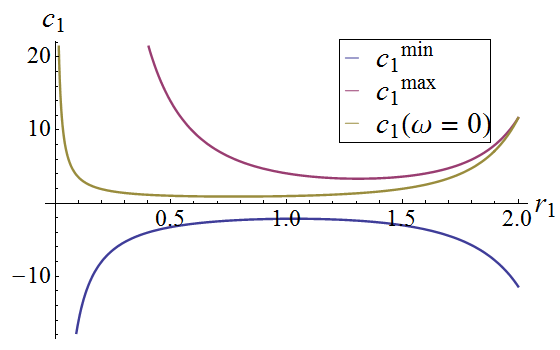} 
\caption{To have energy extraction for small black holes, $\omega^{(1)}$, $c_2$ and $c_1$ must lie between the top and bottom curves in each plot. The middle curve corresponds to $\omega=0$. Note that $\omega^{(1)}_\text{min/max}$ and $c_2^\text{min/max}$ correspond to $c_1^\text{max/min}$.}\label{omega_c_b_ranges}
\end{figure}

Although we are primarily concerned with large black holes in this section, we note more generally that the condition for energy extraction constrains $\omega$ (and thus $c_2$) to the range $\Omega_\infty<\omega<\Omega_H$ (or equivalently $-\frac{(r_1-2)^2+4}{(r_1-2)^2r_1^2}<c_2<\frac{4-r_1}{(r_1-2)^2r_1}$).  Correspondingly, we find constraints on $c_1^\text{min}(r_1)<c_1<c_1^\text{max}(r_1)$, 
which we plot in Fig. \ref{omega_c_b_ranges}. Of course, $c_1$ only has a holographic interpretation in terms of the dual current for large black holes with $r_H > l$.

\subsection{Stability}

As discussed in Sections~\ref{gen} and \ref{setups}, and briefly reviewed again here, Kerr and Kerr-AdS geometries have important differences concerning timelike Killing vectors and ergospheres. For Kerr black holes in asymptotically flat space, there is a unique normalized Killing vector which is timelike at infinity, $K_{\Omega=0}^\mu = \xi_{(t)}^\mu$. This Killing vector becomes spacelike inside the ergosphere, allowing for the possibility of energy extraction from the black hole via super-radiance or the BZ process. In AdS, super-radiant modes would be reflected back off the boundary, leading to a genuine instability. However, as reviewed above, this situation changes in large Kerr-AdS geometries, where a family of Killing vectors remain timelike at infinity. Among this class, the horizon generator $K_{\Om_H}^\mu=\xi_{(t)}^\mu + \Om_H \xi_{(\varphi)}^\mu$ is globally timelike, becoming null on the horizon itself. Thus, there is no ergoregion for the energy flux vector defined by ${\cal T}^\mu = - T^{\mu}_{\nu} K_{\Om_H}^\nu$, which is itself timelike outside the horizon if the dominant energy condition is satisfied.  Hawking and Reall have argued that the existence of this global timelike Killing vector, along with the DEC, ensures stability of Kerr-AdS if $\Om < 1/l$ \cite{HawkingReall}.This argument is apparent in the discussion of section 2, and implies that energy cannot be extracted from large Kerr-AdS black holes by super-radiance or indeed by the BZ process. This argument of course breaks down for small black holes, which then behave in a similar manner to Kerr geometries. 

This stability argument for large black holes ultimately appears consistent with the above conclusions that the boundary rotating fluid is neutral, albeit to low order in the rotation parameter. 
Nonetheless, an important caveat is that the dominant energy condition (DEC) needs to be satisfied. In AdS, violating the DEC is not as dramatic as it would be in flat space since the Breitenlohner-Freedman (BF) bound allows a range of negative mass e.g. for scalar fields. This loophole was noted by Gubser \& Mitra \cite{gm} as a way to realize a Gregory-LaFlamme-type instability for large black holes.
In the present case, it is not clear that the currents which produce a force-free magnetosphere satisfy the DEC,\footnote{The DEC can be verified for the force-free electromagnetic field configuration to ${\cal O}(a^2)$. We find that $-T^{\mu}_{\nu} K^\nu_{\Om_H}$ is indeed a future-directed timelike vector, with the norm scaling as $1/r^2$ as $r\rightarrow \infty$. The ${\cal O}(a^2)$ correction is actually negative, but is necessarily subleading in the slow rotation limit.} but we can try to check this by looking at the asymptotics of the bulk current, given in terms of the bulk solution as $J^\mu=F^{\mu\nu}_{\phantom{\mu\nu};\nu}$. For the boundary field theory directions $\{t,\theta,\varphi\}$, we find the covariant current components,
\begin{align}
J_t & 
          \sim \frac{2Cau (l^2\om^{(1)}+1)}{l^2r^2} + {\cal O}(r^{-5}),\\
J_u & \propto \partial_rB_T(u)=0,\\
J_\varphi & 
   \sim \frac{2Ca^2u(1-u^2)(r_1^2+l^2)(l^2-r_1^2-2l^2r_1^2\om^{(1)})}{r^2l^2r_1^4} +{\cal O}(r^{-4}).
\end{align}
We would like to interpret these falloff conditions in terms of the conformal dimension $\De$ of the dual vector operator according to the boundary coupling ${\cal O}_\mu J^\mu$, and compare with the BF bound. To do this, we can consider modelling the current with a specific bulk field, and for simplicity consider the case of a charged scalar $\ph$ in the bulk, with the current $J_\mu = \phi^*\!\!\stackrel{\leftrightarrow}{D}_\mu\!\!\phi$. Then $J_t = i\phi^*\ph A_t$, and $J_\varphi = i\phi^*\ph A_\varphi$. Recall that the BF bound for a scalar field in 3+1D is $m^2l^2 > -9/3$, where for scalars $\Delta (d-\Delta) = -m^2l^2$. The falloff conditions for the (covariant) current components $J_t \sim J_\varphi \sim 1/r^2 +{\cal O}(1/r^4)$ then imply $\Delta = 1$, i.e. $m^2= -2/l^2$ which is above the BF bound. This allows for two possible normalizable falloff conditions, $\phi \sim \al_1/r + \al_2/r^2+\cdots$, and those above suggest $\al_1\neq 0$ and $\al_2=0$, which is a consistent choice.\footnote{Note that for near-extremal black holes, this dual operator dimension is also in the range for which condensation is possible producing a holographic superfluid.} The result is of course consistent with the stability of the solution. 

The radial component of the current is given by 
\be
 J_r \sim 2Cl^2 au \frac{\om^{(1)} r_1^2 - 1}{r^4r_1^2} + {\cal O}(r^{-6}),
\ee
which in principle sources another scalar operator, independent of ${\cal O}_\mu$. Indeed, expressing $D_\mu$ in BL coordinates, the simple scalar model above would imply $J_r \sim i \phi^*\ptl_r \phi \sim 1/r^3$, which is not consistent with the $1/r^4$ scaling above, suggesting instead a higher dimensional operator. 

To conclude this section, we note that the azimuthal current $J^\varphi$ can change sign from the horizon to the boundary: in $m=1$ units and with $z=\frac{r_1}{r}$,
$J^\varphi_\text{min}=J^\varphi(z=0)\sim1-r_1^2\omega^{(1)}-2r_1\omega^{(1)}$ while 
$J^\varphi_\text{max}=J^\varphi(z=1)\sim1-r_1^2\omega^{(1)}$.
For example, with $\omega^{(1)}=\Omega_H/2=1/(2r_1^2)$, $J^\varphi_\text{min}=1/2- 1/r_1 <0$ while $J^\varphi_\text{max}=1/2$ and the sign change happens closer to the horizon for smaller $r_1$ (i.e., larger black holes). This is consistent with the sign change in the ZAMO frame, suggesting a dominant effect of frame dragging. However, the sign change actually persists for $J^\Phi=J^\varphi+\frac{a}{l^2}J^t$ as measured with respect to the non-rotating frame at infinity.

Another (perhaps related) observation about the current is that in the Kerr case it is everywhere space-like (i.e. `magnetostatic') outside the horizon, while for Kerr-AdS $J_\mu J^\mu$ can change sign. In the presence of both positive and negative charges, a spacelike 'magnetostatic' current is perfectly physical, and indeed is likely the most stable configuration under the assumption that local electric fields are fully screened. An example from our solution is the current $J^\mu(\omega=\Omega_H)\sim\xi_{(t)}^\mu+\Omega_H\xi_{(\varphi)}^\mu$ which satisfies the force-free condition by noting $F_{\mu\nu}\bigl[\xi_{(t)}^\nu+\omega\xi_{(\varphi)}^\nu\bigr]\equiv0$ and can certainly be space-like for small black holes from the previous discussions. The question of whether timelike (i.e. electrically dominant) current domains actually imply instabilities of some sort deserves further investigation.

\section{Concluding Remarks}\label{conclusion}

We have obtained solutions for the force-free BZ process (split monopole magnetosphere) in the slowly rotating Kerr-AdS background up to $\mathcal{O}(a^2)$. The field configuration is poloidal, and can be specified by $A_\varphi$, and thus the main differential equation to solve is that for $f(r)$ (or $f(z)$), the radial $\mathcal{O}(a^2)$ component of $A_\varphi$. In distinction to  the Kerr case originally studied by Blandford and Znajek, the field angular velocity $\omega$, as a parameter in the equation, is not uniquely determined solely from the (minimal) boundary conditions. However, it is  directly related to the toroidal magnetic field $B_T$ due to the horizon regularity constraint for $B^\varphi$ in Kerr-Schild coordinates. Further constraints on $\om$ do arise on imposing specific matching conditions at large radius with a rotating monopole in AdS. However, unlike the asymptotically flat space, these matching conditions are in turn non-unique due to the fact that the AdS boundary is only defined up to a conformal factor, and this allows a class of higher order multipole corrections. Matching to this full class of solutions re-introduces the freedom to vary the field angular velocity. 

The energy-momentum fluxes $T_{t(\varphi)}^\mu$ and bulk currents $J^\mu$ depend, at leading order, on $\omega$, and the `energy' flux into the black hole is given by $\delta E\propto(\omega-\Omega)(\omega-\Omega_H)$ if we use $K_\Omega^\mu=\xi_{(t)}^\mu+\Omega\xi_{(\varphi)}^\mu$ as the energy-defining Killing vector. As expected, unless $\Omega=\Omega_H$ for which the Killing vector is globally time-like, there is an outgoing radial `energy' flux via the BZ process (when $\Omega<\omega<\Omega_H$). The flux is ingoing in the special case of $\Omega = \Omega_H$, i.e. $\delta E\propto(\omega-\Omega_H)^2$,  consistent with the general argument against `energy' extraction with the time coordinate defined according to this specific timelike Killing vector. The globally timelike nature of $K_{\Omega_H}^\mu$ is guaranteed for $r_H>l\geq\sqrt{al}$, i.e., for large black holes. Our small `$a$' expansion was general enough to cover large, intermediate and small black hole regimes. This allowed us to trace the solution into the large black hole regime, and consider the holographic dual description in terms of a thermal fluid rotating in a sphere. The existence of a range of angular velocities for the field seems to correspond simply to the fact that the fluid is neutral to the corresponding (albeit lowest) order in $a$, so that the magnetic field can be freely rotated. In effect the boundary configuration is also `force free', although the solution would need to be extended to higher order in `$a$' to test this in a nontrivial manner.

In concluding this section, it is interesting to consider the implications for the membrane paradigm \cite{kpm}, for which the original BZ process provides an archetypal example \cite{mem1}. The AdS/CFT correspondence generally provides a more explicit holographic translation of the dynamics of black hole horizons, and may assist in providing a more quantitative model of force-free magnetospheres. The membrane paradigm describes the black hole (i.e. the ergosphere) as a `battery' driving the system, within a circuit model for the magnetosphere. Of course the distinctive features of the BZ process in the large black hole regime discussed here make it difficult to provide a more precise AdS/CFT description of the energy extraction process. Nonetheless, this question was one of the original motivations for this work, and it would be interesting to see if the dual field theory picture can be developed further, perhaps by extending the solutions described here away from the slow rotation limit.

\section*{Acknowledgements}

We would like to thank Werner Israel and Pavel Kovtun for helpful discussions.  The work of  X.W. and A.R. is supported 
in part by NSERC, Canada.

\appendix

\section{Force-free solutions in Kerr-Schild coordinates}\label{Appdx_KS_results}
\renewcommand{\theequation}{\thesection.\arabic{equation}}

The analysis in this paper can also be carried out using Kerr-Schild (KS) coordinates (specified with a tilde) which use an alternate foliation $\Sigma_{\tilde t}$ that is horizon penetrating. The metric in this coordinate system is given by \cite{general_Kerr-AdS_KS},
\begin{multline}
{\operatorname d\tilde s}^2=\frac{\Sigma}{\bigl(1+\frac{r^2}{l^2}\bigr)(r^2+a^2)}{\operatorname dr}^2+\frac{\Sigma}{\Delta_\theta}{\operatorname d\theta}^2+\frac{r^2+a^2}{\Xi}\sin^2\theta(\operatorname d\tilde\varphi+\frac{a}{l^2}\operatorname d\tilde t)^2-\frac{\bigl(1+\frac{r^2}{l^2}\bigr)\Delta_\theta}{\Xi}{\operatorname dt}^2\\
+\frac{2mr}{\Sigma}\Bigl[\frac{\Sigma}{\bigl(1+\frac{r^2}{l^2}\bigr)(r^2+a^2)}\operatorname dr-\frac{a}{\Xi}\sin^2\theta\operatorname d\tilde\varphi+\frac{\Delta_\theta}{\Xi}\operatorname d\tilde t\Bigr]^2,
\end{multline}
where $r$ \& $\theta$ are the same as in BL coordinates and $\tilde\varphi$ \& $\tilde t$ are related to the BL coordinates by the transformation 
\cite{general_Kerr-AdS_KS},\footnote{The KS coordinate $\phi$ in \cite{general_Kerr-AdS_KS} is in fact $\phi=\tilde\varphi+\frac{a}{l^2}t$, associated with the non-rotating frame at infinity. Note that \eqref{KS_BL_ph} does not reduce to the coordinate transformation used in \cite{McKGam} in the Kerr limit.}
\begin{align}
\operatorname{d}\tilde\varphi&=\operatorname{d}\varphi+\frac{a\Xi}{\Delta_r\bigl(1+\frac{\Delta_r}{2mr}\bigr)}\operatorname{d}r=\operatorname{d}\varphi+\frac{2mar\Xi}{\Delta_r(r^2+a^2)\bigl(1+\frac{r^2}{l^2}\bigr)}\operatorname{d}r\label{KS_BL_ph}\\
\operatorname{d}\tilde t&=\operatorname{d}t+\frac{2mr}{\Delta_r\bigl(1+\frac{r^2}{l^2}\bigr)}\operatorname{d}r\label{KS_BL_t}.
\end{align}
Note that
\begin{equation}
\frac{\tilde\varphi_{,r}}{\tilde t_{,r}}=\frac{a\Xi}{(r^2+a^2)}\overset{r=r_H}{=\joinrel=\joinrel=}\Omega_H,
\end{equation}
and thus the transformation \{$r,\theta,\tilde\varphi(\varphi,r),\tilde t(t,r)$\} does not affect ZAMO 4-velocity components, i.e., $\tilde K_{\Omega_B}^{\tilde\mu}=K_{\Omega_B}^\mu$ (as long as $K_{\Omega_B}^r=0$) and
\begin{align}
\Omega_B&=-\tilde{\beta}^\varphi-\frac{\tilde{h}_{r\varphi}}{\tilde{h}_{\varphi\varphi}}\tilde{\beta}^r\label{OmegaB_KS},\\
\begin{split}
K_{\Omega_B}^\mu{K_{\Omega_B}}_\mu&=-\tilde{\alpha}^2+\frac{\tilde{h}_{rr}\tilde{h}_{\varphi\varphi}-\tilde{h}_{r\varphi}^2}{\tilde{h}_{\varphi\varphi}}(\tilde{\beta}^r)^2\label{ZAM_Ksq_KS},\\
&\overset{r=r_H}{=\joinrel=\joinrel=}0\ne\tilde{\alpha}^2\tilde{n}_\mu\tilde{n}^\mu=-\tilde{\alpha}^2.
\end{split}
\end{align}
Since $K_{\Omega_B}^i\ne-\tilde{\beta}^i$ and $K_{\Omega_B}^\mu$ is not parallel to $\tilde{\alpha}\tilde n^\mu$, the ZAMO is no longer a fiducial observer. On the horizon where $K_{\Omega_B}^\mu$ is the outgoing null generator, $\tilde{\alpha}\tilde n^\mu$ lies inside the light cone and is ingoing. The horizon condition \eqref{ZAM_Ksq_KS} does not make any metric component singular. It is worth noting that for both BL and KS coordinates, $g^{rr}=0$ on the horizon which is a null constant-$r$ hypersurface.

The transformation only affects the contravariant $\varphi,t$ and covariant $r$ components of tensorial objects ($g_{\mu\nu}, F_{\mu\nu}, dT_\mu,\dotsc$); in particular, the following quantities are all invariant: functions of $(r,\theta)$ (e.g., $\det g_{\mu\nu}$), the derivatives $\partial_\mu$ (hence the conditions $(\ldots)_{,t}=0=(\ldots)_{,\varphi}$ and the bracket structure \eqref{brk}) and the definition and value of $\omega$. $B^\varphi$ and $B_T$ transform as
\begin{align}
B^\varphi&=\frac{F_{ru}}{\sqrt{-g}}=\frac{\tilde F_{ru}+\tilde\varphi_{,r}\tilde F_{\varphi u}+\tilde t_{,r}\overbrace{\tilde F_{tu}}^{=-\omega\tilde F_{\varphi u}}}{\sqrt{-g}}\\
&=\tilde B^\varphi+(\omega\tilde t_{,r}-\tilde\varphi_{,r})\tilde B^r,\\
B_T&=(g_{\varphi\varphi}g_{tt}-g_{\varphi t}^2)B^\varphi\\
&=\tilde B_T+(g_{\varphi\varphi}g_{tt}-g_{\varphi t}^2)(\omega\tilde t_{,r}-\tilde\varphi_{,r})\tilde B^r.\label{BT_BL_KS}
\end{align}
$B_T$ is $r$-independent and $\tilde B_T\sim\Delta_r\tilde B^\varphi=0$ on the horizon (for regular $\tilde B^\varphi$), so
\begin{align}
B_T&=(g_{\varphi\varphi}g_{tt}-g_{\varphi t}^2)(\omega\tilde t_{,r}-\tilde\varphi_{,r})\tilde B^r\big\rvert_{r=r_H}\\
&=-\frac{\Delta_\theta(1-u^2)}{\Xi}\frac{r_H^2+a^2}{r_H^2+a^2u^2}(\omega-\Omega_H)A_{\varphi,u}\\
&=-(1-u^2)(\omega-\Omega_H)A_{\varphi,u}+\mathcal{O}(a^3)\label{BT_omega_A_phi_a}.
\end{align}
Note that $B_T$'s defined using $u$ and $\theta$ differ by a  sign, and $\sqrt{-g(\theta)}=\frac{\Sigma}{\Xi}\sin\theta,\sqrt{-g(u)}=\frac{\Sigma}{\Xi}$.

Using the same ansatz \eqref{ansatz_A_phi}--\eqref{ansatz_BT} for \{$A_\varphi,\omega,\tilde B_T$\} in the small `$a$' expansion we find
\begin{align}
\widetilde{dT}_\varphi&=aC^2\biggl[\frac{\tilde B_{T_{,r}}^{(1)}}{Cr^2}+2ml^2(1-u^2)\frac{r^2(3r^2+l^2)\omega^{(1)}-5r^2-3l^2}{r^6(r^2+l^2)^2}\biggr]\label{dT_ph_KS},\\
\widetilde{dT}_t&=-\omega \widetilde{dT}_\varphi\label{dT_t_KS},\\
\widetilde{dT}_r&=\frac{\tilde B_Tr^2}{C(1-u^2)\Delta_0} \widetilde{dT}_\varphi=\frac{\tilde B^\varphi r^2}{C} \widetilde{dT}_\varphi\label{dT_r_KS},\\
\widetilde{dT}_u&=(\text{expression involving 2nd derivatives of }A_\varphi)\label{dT_u_KS},
\end{align}
where $\omega$ is constant. $\widetilde{dT}_\varphi$, $\widetilde{dT}_t$ and $\widetilde{dT}_u$ are the same as in BL coordinates; $\widetilde{dT}_r$ is now directly proportional to $\widetilde{dT}_\varphi$. The two independent equations are \eqref{dT_ph_KS}=0 \& \eqref{dT_u_KS}=0. Solving $\eqref{dT_ph_KS}=0$ for $\tilde B_T^{(1)}$ and imposing $\tilde B_T^{(1)}(r=r_1)=0$ yields
\begin{equation}\label{BT_omega_KS}
\tilde B_T^{(1)}=-2Cml^2(1-u^2)\left[\frac{\omega^{(1)}-\frac{1}{r^2}}{r(r^2+l^2)}-\frac{\omega^{(1)}-\frac{1}{r_1^2}}{r_1(r_1^2+l^2)}\right].
\end{equation}
Substitution into \eqref{BT_BL_KS} leads to
\begin{equation}\label{BT_omega}
B_T=C(1-u^2)\Bigl(\omega-\frac{a}{r_1^2}\Bigr)+\mathcal{O}(a^3),
\end{equation}
which agrees with \eqref{BT_omega_A_phi_a}. This fixes the sign ambiguity of $B_T^c$ in \eqref{BTsq_omega}.

\section{Analytic force-free magnetosphere for small Kerr-AdS black holes} \label{AppB}
\renewcommand{\theequation}{\thesection.\arabic{equation}}

In this appendix, we determine an analytic solution for the force-free magnetosphere about a `small' Kerr-AdS black hole.
More precisely, in the slow rotation, small `$a$', limit we also expand in $r_H/l$ and consider the leading correction of ${\cal O}(r_H^2/l^2)$. For simplicity below, we will only keep track of the order in $1/l$, and refer to this as the $1/l$ expansion. 

Starting with the ansatz
\begin{align}
A_\varphi^{(2)}(r,u)&=\bigl[f_{[0]}(r)+f_{[2]}(r)l^{-2}\bigr]Cu(1-u^2),\\
\omega^{(1)}&=\frac{1}{8m^2}+\omega_{[2]}^{(1)}l^{-2},\\
\frac{B_T^{(1)}}{C}&=\omega^{(1)}-\frac{1}{r_1^2}=-\frac{1}{8m^2}+(\omega_{[2]}^{(1)}-2)l^{-2},
\end{align}
where $\omega_{[2]}^{(1)}$ is constant, the equation $dT_u=0$ \eqref{dT_u} has the expansion $dT_\theta^{[0]}+dT_\theta^{[2]}l^{-2} + \cdots=0$. The leading order equation $dT_\theta^{[0]}=0$ is solved by the known BZ solution in the Kerr geometry:
\begin{multline}\label{0th_fr_sln}
f_{[0]}(r)=\frac{2r-3m}{m^4}\left[\frac{r^2}{16m}\left(2\mathrm{dilog}\frac r{2m}+\ln^2\frac r{2m}+\frac{\pi^2}3\right)-\frac r4-\frac m8-\frac{m^2}{9r}\right]\\
-\frac{1}{m^4}\left(\frac{r^2}2-\frac{mr}4-\frac{m^2}{12}\right)\ln\frac r{2m}.
\end{multline}
The next-to-leading-order equation $dT_\theta^{[2]}=0$ has a solution of the form
\begin{equation}\label{fx_formal}
f_{[2]}(x)=\frac{1}{36}\bigl[3C_1h_1(x)+C_2h_2(x)+h_2(x)\int h_1(x)h_3(x)\mathrm dx-h_1(x)\int h_2(x)h_3(x)\mathrm dx\bigr],
\end{equation}
where $x\equiv\frac{r}{2m}$ and
\begin{align}
h_1(x)&=-4x^2(4x-3),\\
h_2(x)&=-12x^2(4x-3)\ln\Big(1-\frac1x\Big)-2(24x^2-6x-1),
\end{align}
\begin{multline}
h_3(x)=-\frac92x^2(8x-3)(2\mathrm{dilog}x+\ln^2x)+\frac{3x(48x^3-90x^2+43x-2)\ln x}{2(x-1)^2}-\\
\frac{48\pi^2x^6-(66\pi^2+288)x^5+(14\pi^2+324)x^4-(35+12\omega_{[2]}^{(1)})x^3-7x^2+12\omega_{[2]}^{(1)}}{4x^2(x-1)}.
\end{multline}
The function $h_3(x)$ can be written in a slightly different form 
\begin{multline}
h_3(x)=-9x^2(8x-3)\Big(\overbrace{\mathrm{Li}_2\frac1x-\ln(x-1)\ln x+\ln^2x}^{=\mathrm{Li}_2\frac1x+\mathrm{Li}_1\frac1x\ln x}\Big)+\frac{3x(48x^3-90x^2+43x-2)\ln x}{2(x-1)^2}\\
+\frac{288x^5-324x^4+(35+12\omega_{[2]}^{(1)})x^3+7x^2-12\omega_{[2]}^{(1)}}{4x^2(x-1)},
\end{multline}
using the dilogarithm identities
\begin{equation}\label{di_poly}\begin{split}
\mathrm{dilog}x=\mathrm{Li}_2(1-x)&=\mathrm{Li}_2\frac1x+\frac12\ln x\ln\frac x{(x-1)^2}-\frac{\pi^2}6\\
&=\mathrm{Li}_2\frac1x-\ln(x-1)\ln x+\frac12\ln^2x-\frac{\pi^2}6\\
&=\mathrm{Li}_2\frac1x+\mathrm{Li}_1\frac1x\ln x-\frac12\ln^2x-\frac{\pi^2}6,\qquad (x>1)
\end{split}\end{equation}

With some manipulation, the integrals in \eqref{fx_formal} could be evaluated using \emph{Maple}. In particular, the second integral can be simplified using integration by parts, as sketched below,
\begin{equation}\label{parts}
\int h_2(x)h_3(x)\mathrm dx=\int\mathrm{part}_1\mathrm dx+\underbrace{\int\ln\Big(\frac{x-1}x\Big)\mathrm{part}_2\mathrm dx}_{=\mathrm{part}_{1a}+\big(\int\mathrm{part}_{1b}\mathrm dx+\underbrace{\int\ln^2x\mathrm{part}_{2b}\mathrm dx}_{\text{integration by parts}}\big)}.
\end{equation}
Before presenting the explicit result, we note that certain terms in \eqref{parts} appear to be complex. For example,
\begin{equation}
\frac{4212}{35}\left[\mathrm{Li}_3x-\mathrm{Li}_2x\ln x-\frac12\ln(1-x)\ln^2x\right],
\end{equation}
since it contains $\mathrm{Li}_sx$, is only real for $x\leq1$ ($\ln(1-x)=-\mathrm{Li}_1x$) while our $x$ is in $[1,\infty)$. However, all the imaginary parts actually cancel out as can be shown using the following identities
\begin{align*}
\mathrm{Li}_1x&=\mathrm{Li}_1\frac1x-\ln x-i\pi\quad\big(\Leftrightarrow\ln(1-x)=\ln(x-1)+i\pi\big),\\
\mathrm{Li}_2x&=-\mathrm{Li}_2\frac1x-\frac12\ln^2x+\frac{\pi^2}3-i\pi\ln x\qquad\qquad\qquad\qquad\qquad\qquad(x>1),\\
\mathrm{Li}_3x&=\mathrm{Li}_3\frac1x-\frac16\ln^3x+\frac13\pi^2\ln x-\frac12i\pi\ln^2x,
\end{align*}
or more generally
\begin{equation}
\mathrm{Li}_sx+(-1)^s\mathrm{Li}_s\frac1x=2\sum^{\lfloor s/2\rfloor}_{k=0}\frac{\ln(-x)^{(s-2k)}\mathrm{Li}_{2k}(-1)}{(s-2k)!}.
\end{equation}

The final solution is given by
\begin{multline}
f_{[2]}(x)=\frac{78x^2(4x-3)}{35}\left(6\mathrm{Li}_3\frac1x+4\mathrm{Li}_2\frac1x\ln x+\mathrm{Li}_1\frac1x\ln^2x\right)\\
-\frac{2(120x^5+195x^4-234x^3-312x^2+78x+13)}{35}\left(\mathrm{Li}_2\frac1x+\mathrm{Li}_1\frac1x\ln x\right)\\
-\frac{13(24x^2-6x-1)\ln^2x}{35}+\frac{x(240x^4+270x^3-1951x^2+1397x-26)\ln x}{35(x-1)}\\
+\frac{48x^4}7+\frac{90x^3}{7}-\frac{2659x^2}{42}+\frac{1427x}{105},
\end{multline}
where we have set $\omega_{[2]}^{(1)}=1/2$ to remove the $\mathcal{O}(x^3,x^2)$ divergences at large $x$, while $\mathcal{O}(x)$ and $\ln x$ divergences remain (implying finite field strengths); in the two limits,
\begin{align}
f_{[2]}(x\rightarrow\infty)&=-\frac x3+\frac{\ln x}{30}-\frac{833}{1800},\\
f_{[2]}(x=1)&=\frac{468\zeta(3)}{35}+\frac{4\pi^2}3-\frac{9434}{315}.
\end{align}
Thus we have
\begin{align}
\omega^{(1)}&=\frac{1}{8m^2}+\frac{1}{2l^2}<\frac{\Omega_H}{a}=\frac{1}{4m^2}+\frac{2}{l^2},\\
B_T&=-C(1-u^2)\bigl(\frac{1}{8m^2}+\frac{3}{2l^2}\bigr),
\end{align}
consistent with an outgoing energy flux in this `small' black hole limit.

\section{Force-free equations in the Newman-Penrose (NP) formalism}\label{AppC}

For completeness, in this Appendix we will also rewrite the Kerr-AdS force-free equations $F_{\mu\nu}J^\nu=0$ in first-order form using the NP formalism \cite{np}. The NP variables for the electromagnetic field are conveniently expressed as the coefficients in an expansion of the anti-self-dual part of $F_{\alpha\beta}$:
\begin{equation}\label{Fasd_UWV}
F^-_{\alpha\beta}=\phi_0U_{\alpha\beta}+\phi_1W_{\alpha\beta}+\phi_2V_{\alpha\beta},
\end{equation}
where $F^-_{\al\beta} = F_{\alpha\beta}+ i {^\star}\!F_{\alpha\beta}$ and the basis for anti-self-dual bi-vectors is formed using the NP null tetrad $\{ l,n,m,\bar{m}\}$,
\begin{equation}
U_{\alpha\beta}\equiv 2\bar m_{[\alpha}n_{\beta]},\quad W_{\alpha\beta}\equiv 2(n_{[\alpha}l_{\beta]}+m_{[\alpha}\bar m_{\beta]}),\quad V_{\alpha\beta}\equiv 2l_{[\alpha}m_{\beta]}.
\end{equation}
The tetrad indices $\{(1),(2),(3),(4)\}$ correspond to contractions with $\{ l,n,m,\bar{m}\}$ respectively, e.g., $F_{(1)(3)}=F_{\alpha\beta}l^\alpha m^\beta$, etc..

As noted in the main text, only two of the $F^{(a)(b)}J_{(b)}=0$ equations are independent which we choose to be
\begin{align}
\phi_0J_{(2)}&=\bar\phi_2J_{(1)}+(\phi_1-\bar\phi_1)J_{(3)},\label{FJ2}\\
\phi_0J_{(4)}&=(\phi_1+\bar\phi_1)J_{(1)}-\bar\phi_0J_{(3)}.\label{FJ4}
\end{align}
Here the $J_{(a)}$ contain terms with different NP variables acted on by directional derivatives (along tetrad vectors) and multiplied by various spin coefficients. One can achieve more compact forms by introducing the following quantities (specifying now to the Kerr-AdS metric with a Kinnersley-like tetrad in BL coordinates \cite{Kerr-AdS_tetrad})
\begin{equation}
\Phi_1\equiv\Delta_r\rho\phi_0+\frac{2\phi_2}{\rho},\qquad \Phi_2\equiv\Delta_r\rho\phi_0-\frac{2\phi_2}{\rho},
\end{equation}
and
\begin{align}
2J_{(2)}^T&\equiv J_{(2)}+\frac{J_{(1)}}{k_1}, & 2J_{(2)}^P&\equiv J_{(2)}-\frac{J_{(1)}}{k_1},\\
2J_{(4)}^T&\equiv J_{(4)}+\frac{J_{(3)}}{k_2}, & 2J_{(4)}^P&\equiv J_{(4)}-\frac{J_{(3)}}{k_2},
\end{align}
where
\begin{equation}
\rho=-\frac{1}{r-ia\cos\theta},\quad k_1\equiv\frac{2\Sigma}{\Delta_r},\quad k_2\equiv-\frac{\bar\rho}{\rho}.
\end{equation}
The superscript ``$T/P$'' indicates that, e.g., $J_{(2)}^{T/P}$ only involves the contractions of $J_\mu$ with the toroidal/poloidal components of $n^\mu$.\footnote{Note that in the Kerr limit, $\Phi_1$ \& $\Phi_2$ are proportional to the electric and magnetic field components in the orthonormal frame associated with the Carter tetrad, e.g., $\Phi_1\sim(E_3+iB_3)_\text{Carter}$ etc., and $J^{T/P}$ are proportional to components of $J_\mu$ in this orthonormal frame \cite{Carter_tetrad}.}

$J^{T/P}$ only involve simple derivatives of the new field quantities:
\begin{align}
J_{(2)}^P&=-\dfrac{\partial_\theta(\Phi_1\sin\theta\sqrt{\Delta_\theta})}{4\sqrt 2\Sigma\sin\theta}, & J_{(4)}^P&=\dfrac{\rho\partial_r(\Phi_1)}{4},\\
J_{(2)}^T&=\dfrac{\partial_\theta(\Phi_2\sin\theta\sqrt{\Delta_\theta})}{4\sqrt 2\Sigma\sin\theta}+\dfrac{\Delta_r\rho^2}{2\Sigma}\partial_r\dfrac{\phi_1}{\rho^2} & J_{(4)}^T, &=-\dfrac{\rho\partial_r(\Phi_2)}{4}+\frac{\sqrt{\Delta_\theta}\rho^3}{\sqrt2}\partial_\theta\dfrac{\phi_1}{\rho^2}.
\end{align}
The condition $F_{\varphi t}=0$ assumed under stationarity and axisymmetry becomes
\begin{equation}\label{SAS}
\sqrt2F_{\varphi t}=\frac{\sin\theta\sqrt{\Delta_\theta}}{\Xi}\Im\Phi_1=0,
\end{equation}
which can also be inferred from the reality of $J_{(2)}^P$. The reality of $J_{(2)}^T$ imposes another constraint,
\begin{equation}\label{reality_J2T}
0=\Im J_{(2)}^T=\bigl[\partial_r(\Sigma\Im\phi_1)+2a\cos\theta\Re\phi_1\bigr]\frac{\Delta_r}{2\Sigma^2}+\frac{\partial_\theta(\sin\theta\sqrt{\Delta_\theta}\Im\Phi_2)}{4\sqrt2\sin\theta\Sigma}.
\end{equation}
The force-free condition implies the degeneracy of the electromagnetic field,
\begin{align}
&\det F_{(a)(b)}=[F_{(m)(n)}{^\star}F^{(m)(n)}]^2/16=0\\
\quad\Leftrightarrow\quad&\Im(\phi_0\phi_2-\phi_1^2)=\Im\Bigl(\frac{\Phi_1^2-\Phi_2^2}{8\Delta_r}-\phi_1^2\Bigr)=0,\label{ff}
\end{align}
which combined with \eqref{SAS} gives
\begin{equation}\label{ffSAS}
-8\Delta_r\Re\phi_1\Im\phi_1=\Re\Phi_2\Im\Phi_2\quad\Leftrightarrow\quad\Im(\Phi_2^2+8\Delta_r\phi_1^2)=0.
\end{equation}
The relations \eqref{SAS} \& \eqref{ffSAS} will be implicitly assumed and applied in the following discussion.

Now we can rewrite \eqref{FJ2} \& \eqref{FJ4} in terms of the new quantities. One combination yields
\begin{equation}\label{FJ_new1}
\Im\Phi_2J_{(2)}^P-2\bar\rho\Delta_r\Im\phi_1J_{(4)}^P=0\quad\Leftrightarrow\quad(2\sqrt2\Delta_r\Im\phi_1\partial_r+\sqrt{\Delta_\theta}\Im\Phi_2\partial_\theta)(\Phi_1\sin\theta\sqrt{\Delta_\theta})=0.
\end{equation}
Then \eqref{FJ2} \& \eqref{FJ4} reduce to
\begin{equation}
\Re\Phi_2J_{(2)}^T+2i\bar\rho\Delta_r\Im\phi_1J_{(4)}^T+\Re\Phi_1J_{(2)}^P=0\quad\Leftrightarrow\quad
\end{equation}
\begin{multline}\label{FJ_new2}
\Big(\frac{\Re\Phi_2}{2\sqrt2\Delta_r}\partial_\theta-i\frac{\Im\phi_1}{\sqrt{\Delta_\theta}}\partial_r\Big)(\Phi_2\sin\theta\sqrt{\Delta_\theta})+\rho^2\sin\theta(\Re\Phi_2\partial_r+i2\sqrt2\sqrt{\Delta_\theta}\Im\phi_1\partial_\theta)\frac{\phi_1}{\rho^2}\\
-\frac{\Phi_1\partial_\theta(\Phi_1\sin\theta\sqrt{\Delta_\theta})}{2\sqrt2\Delta_r}=0.
\end{multline}
The two equations \eqref{FJ_new1} \& \eqref{FJ_new2} are the primary force-free equations, equivalent to \eqref{FJ2} \& \eqref{FJ4}.

To make contact with field quantities used in the main text, we have the relations
\begin{align}
\Xi\partial_rA_\varphi&=-2a\sin^2\theta\Re\phi_1-\frac{(r^2+a^2)\sin\theta\sqrt{\Delta_\theta}}{\sqrt2\Delta_r}\Im\Phi_2,\label{A_phi_r_new}\\
\Xi\partial_\theta A_\varphi&=-\frac{a\sin^2\theta}{\sqrt2\sqrt{\Delta_\theta}}\Re\Phi_2+2(r^2+a^2)\sin\theta\Im\phi_1,\label{A_phi_th_new}\\
\sqrt2\Xi B_T&=\sin\theta\sqrt{\Delta_\theta}\Phi_1,\label{BT_new}\\
\frac{\omega}{\Xi}&=\frac{\Re\Phi_2-2\sqrt2a\sin\theta\sqrt{\Delta_\theta}\Im\phi_1}{a\sin^2\theta\Re\Phi_2-2\sqrt2(r^2+a^2)\sin\theta\sqrt{\Delta_\theta}\Im\phi_1},\label{omega_new}
\end{align}
and the energy flux,
\begin{equation}
T_t^r=\frac{\Phi_1}{2\Sigma}(\Re\Phi_2-2\sqrt2a\sin\theta\sqrt{\Delta_\theta}\Im\phi_1).
\end{equation}
It can be checked, using \eqref{A_phi_r_new}--\eqref{omega_new}, that \eqref{FJ_new2} indeed reduces to \eqref{fr_eqn0} for the slowly rotating monopole ansatz. The equations \eqref{FJ_new1} \& \eqref{FJ_new2} form a useful starting point for further study of force-free solutions in Kerr-AdS away from the slow-rotation ansatz.

\bibliography{paper-bib}

\end{document}